%% file: main.tex
\documentclass[letterpaper,11pt]{article}
\usepackage[margin=1in]{geometry}
\usepackage[utf8]{inputenc} 
\usepackage[T1]{fontenc}    
\usepackage[colorlinks=true,breaklinks=true,bookmarks=true,urlcolor=blue,citecolor=blue,linkcolor=blue,bookmarksopen=false,draft=false]{hyperref}
\usepackage{url}            
\usepackage{booktabs}       
\usepackage{multirow}
\usepackage{microtype}      
\usepackage{amsfonts,amsmath, amsthm,mathtools,stmaryrd}
\usepackage{xcolor}
\usepackage[inline]{enumitem}
\usepackage{appendix}
\usepackage{float, graphicx, subcaption}
\usepackage{bbold}
\usepackage{natbib}
\usepackage[mathscr]{euscript}
\usepackage{tikz}
\renewcommand{\cite}[1]{\citep{#1}}

\newtheorem{remark}{Remark}[section]

\newtheorem{proposition}{Proposition}[section]

\def\argmin_#1{\underset{#1}{\mathrm{arg\,min\, }}}
\def\argmax_#1{\underset{#1}{\mathrm{arg\,max\, }}}

\makeatletter
\def\dasharrowfill@#1#2#3#4{%
        $\m@th
        \thickmuskip0mu
        \medmuskip\thickmuskip
        \thinmuskip\thickmuskip
        \relax
        #4#1\mkern2mu
        \xleaders\hbox{$#4\mkern2mu#2\mkern2mu$}\hfill
        \mkern2mu
        #3$%
}

\def\dashleftarrowfill@{\dasharrowfill@\leftarrow\relbar\relbar}
\def\dashrightarrowfill@{\dasharrowfill@\relbar\relbar\rightarrow}
\def\dashleftrightarrowfill@{\dasharrowfill@\leftarrow\relbar\rightarrow}
\def\dashLeftarrowfill@{\dasharrowfill@\Leftarrow\Relbar\Relbar}
\def\dashRightarrowfill@{\dasharrowfill@\Relbar\Relbar\Rightarrow}
\def\dashLeftrightarrowfill@{\dasharrowfill@\Leftarrow\Relbar\Rightarrow}

\providecommand*\xdashleftarrow[2][]{%
  \ext@arrow 0055{\dashleftarrowfill@}{#1}{#2}}
\providecommand*\xdashrightarrow[2][]{%
  \ext@arrow 0055{\dashrightarrowfill@}{#1}{#2}}
\providecommand*\xdashleftrightarrow[2][]{%
  \ext@arrow 0055{\dashleftrightarrowfill@}{#1}{#2}}
\providecommand*\xdashLeftarrow[2][]{%
  \ext@arrow 0055{\dashLeftarrowfill@}{#1}{#2}}
\providecommand*\xdashRightarrow[2][]{%
  \ext@arrow 0055{\dashRightarrowfill@}{#1}{#2}}
\providecommand*\xdashLeftrightarrow[2][]{%
  \ext@arrow 0055{\dashLeftrightarrowfill@}{#1}{#2}}
\makeatother
\usepackage{nicefrac}       
\usepackage{wrapfig}         
\usepackage{algorithm}
\usepackage{algpseudocode}
\usepackage{caption}
\usepackage{mwe}
\begin{document}
\title{Statistical analysis of team formation and player roles in football}

\author{Ali Baouan \thanks{Centre de Math\'ematiques Appliqu\'ees, Ecole Polytechnique.}}
\date{}
\maketitle

\input{sections/introduction}
\input{sections/methodology}
\input{sections/results}

\input{sections/conclusion}

\clearpage

\bibliographystyle{apalike}
\bibliography{biblio}

\input{sections/appendix}

\end{document}

%% file: sections/introduction.tex
\begin{abstract}
The availability of tracking data in football presents unique opportunities for analyzing team shape and player roles, but leveraging it effectively remains challenging. This difficulty arises from the significant overlap in player positions, which complicates the identification of distinct roles and team formations. In this work, we propose a novel model that incorporates a hidden permutation matrix to simultaneously estimate team formations and assign roles to players at the frame level. To address the cardinality of permutation sets, we develop a statistical procedure to parsimoniously select relevant matrices prior to parameter estimation. Additionally, to capture formation changes during a match, we introduce a latent regime variable, enabling the modeling of dynamic tactical adjustments. This framework disentangles player locations from role-specific positions, providing a clear representation of team structure. We demonstrate the applicability of our approach using player tracking data, showcasing its potential for detailed team and player analysis.
\end{abstract}

\section{Introduction}

Football is inherently a spatial game where teams strategically position players to either advance the ball or restrict the opposition's attack. This involves two primary considerations: determining the areas that should be occupied by the team's eleven and assigning players to these locations. These decisions are guided by the choice of a formation that the manager makes, which defines the relative roles of players and their positions. Football formations have been a cornerstone of the sport, shaping the strategic landscape and influencing the flow of the game. Historically, the evolution of formations reflects the shifting philosophies and tactical innovations that have defined different eras of football. In the early days, formations like the 2-3-5 emphasized offensive play and attacking flair, aligning with the more open and less structured style of play. As the game progressed, the need for greater defensive organization and balance led to the adoption of formations such as the 4-4-2 and 4-3-3, which provided a more structured approach to both offense and defense. In the late 20th and early 21st centuries, even more sophisticated systems emerged, incorporating flexibility and adaptability to counter diverse playing styles and opponents. These advancements in formation strategies and spatial organization have not only enhanced team performance but also elevated the tactical depth of football. For this reason, studying both the formation and the dynamics of role assignment can provide valuable insights into a team's strategy and performance.
\\
\noindent
\\
Automatic detection of formations has emerged as a critical area of interest in sports analytics. This is particularly important because of the availability of tracking data which consists of a collection of frames, where each frame represents a snapshot of the players' locations at a specific timestamp. The ability to identify formations enables researchers and practitioners to analyze team strategies with precision, overcoming the limitations of manual annotation, which is labor-intensive and prone to errors. A natural approach is to consider every player as responsible of a role and then estimate his average position to get an idea of the structure. The primary challenge to this naive approach lies in the significant overlap in player locations during a game, resulting in intersecting roles that give little insight into the strategy of the team. The observed overlap is due to three main reasons. First, the team moves as a block across the pitch, causing all players to cover extensive areas of the field during a game. This effect can be mitigated by centering the frames around the mean position in each frame and scaling by the standard deviation along each coordinate. This ensures we work with relative positions rather than absolute positions. The authors of \cite{narizuka2019clustering} propose an alternative approach by working with the adjacency matrices of the Delaunay graph of each frame. Second, the overlap is also due to the fact that players dynamically swap roles during the game. Several studies highlight the importance of this effect by proposing innovative methodologies to estimate the formation in the presence of player swaps. The authors of \cite{bialkowski2016discovering} present a procedure based on the K-means algorithm for determining roles on the pitch. Their methodology begins with a naive assignment of each player to a role and computes the corresponding center locations and covariance matrices. For each frame, players are optimally reassigned to maximize the total likelihood of each assignment. The role centers and covariance matrices are then updated based on the current assignments, and this process is iterated until convergence. The third source of overlap between players is the change of formation. In particular, teams can alter their structure depending on the phase of play, for example in offense and defense. Some studies explore the task of estimating multiple formations in a single game to account for changes in different phases of play. The authors of \cite{bauer2023putting} introduce a semi-supervised approach that first trains a convolutional neural network to segment games into discrete, non-overlapping phases of play. Then, a formation is identified in each phase of play using the methodology introduced in \cite{shaw2019dynamic}. Collectively, these advancements in automatic formation detection provide a robust foundation for developing metrics to quantify role-swapping rates, strategy shifts, and the effectiveness of tactical adaptations.
\\
\noindent
\\
 In this work, we address the three sources of overlap in order to identify the formation used in a game. Our methodology  simultaneously assigns roles to players  and determines distinct phases of play in a model-based way and at a granular, frame-by-frame level. Our objectives are hence threefold:
 \begin{enumerate}
     \item  Estimating the formation used in a game using tracking data.
     \item Designing both aggregate and instantaneous metrics that quantify the rate of role swapping during a game.
     \item Incorporating in a non-supervised way the possibility of a change of regime.
 \end{enumerate}
 To do so, we build on the work from \cite{bialkowski2016discovering} and propose a probabilistic approach to the estimation of formation and role assignment. We model the player locations as follows: at each frame, $11$ role locations are sampled randomly from associated areas, then these locations are assigned to players through a random permutation. The distributions from which the role locations are sampled define the team's formation, while the probability distribution over permutations captures the links between players and roles. Specifically, we assume that the location of each role follows an independent normal distribution with unknown parameters. Furthermore, since the role assignment permutation is not directly observable, it is treated as a latent variable. This setup corresponds to a special case of Gaussian Mixture Model where the Gaussian components share parameters through permutations. 
 \\
\noindent
\\
Gaussian Mixture Models (GMMs) are a powerful and versatile probabilistic framework widely used for modeling data distributions and uncovering underlying patterns in a variety of fields, including machine learning, computer vision, and statistical analysis. By representing data as a weighted sum of multiple Gaussian components, GMMs capture complex, multi-modal distributions that simpler models often fail to address, see \cite{reynolds2009gaussian}. Their flexibility allows for effective clustering, density estimation, and anomaly detection, making them a popular choice for unsupervised learning tasks, see \cite{bishop2006pattern}. Their interpretability has made GMMs an essential tool in both academic research and practical applications. In our work, the latent variable is the permutation between roles and its correct inference allows us to model and detect the presence of player swaps during a game. It also allows us to produce an unbiased estimate of the team formation. To approximate the parameters of GMMs, the Expectation-Maximization (EM) algorithm is commonly used, iteratively alternating between expectation and maximization steps to efficiently maximize the likelihood function, see \cite{dempster1977maximum}.
 \\
 \noindent
 \\
Our modeling choice of player locations allows for uncertainty in the role assignment at each frame. In fact, conditionally on the observed player locations, it may be difficult to determine whether a swap happened between players if they are close on the pitch. While the K-means based approach in  \cite{bialkowski2016discovering} selects the optimal assignment at each frame, our model-based approach performs a \textit{soft}-assignment of roles, by considering the probability of each player-to-role assignment. Furthermore, recognizing that strategy changes may occur mid-game, we introduce another latent variable to represent the phase of play. The Gaussian mixture setup is well adapted to the addition of this unobservable feature and allows for its non-supervised estimation. Within each state, both the formation and the distribution of likely permutations can change.  
\\
\noindent
\\
Modeling and inferring probability distributions over permutations poses significant computational challenges due to the memory requirements of storing $d! - 1$ weights, which become infeasible even for small $d$. In the work of \cite{bialkowski2016discovering}, this problem is bypassed since it is only necessary to find the optimal permutation for each frame. This can be done using the Hungarian algorithm with a time complexity of $\mathcal{O}(d^3)$, where $d=11$ is the number of players. Several approaches have been proposed to address the difficulty of modeling probability distributions in permutation spaces. For instance, \cite{huang2009fourier} approximates probability distributions using the first few Fourier components, while \cite{plis2011directional} extends the set of permutations into a continuous subspace of $\mathbb{R}^{d \times d}$, using a parametric von Mises-Fisher distribution. In this work, we introduce a novel method tailored to our dataset. With the assumption that only a parsimonious subset of permutations has non-negligible probability, we develop a criterion to discard unlikely permutations. This hypothesis is consistent with the observed dynamics in the game of football. In fact, only a few players switch positions, with some players, such as center-backs and goalkeepers, remaining fixed in their positions.
\\
\noindent
\\
The criterion we define to discard permutations is based on the observation that the locations occupied by players that swap roles show significant levels of overlap. Therefore, absence of overlap is a good signal that the players do not swap roles. To measure the overlap between two distributions using empirical samples, several works have focused on approximating the densities using k-nn estimators or histograms, see \cite{freedman1981histogram,poczos2011estimation}. However, these approaches suffer from the curse of dimensionality and can not be scaled to discard a large number of permutations. As an alternative, we choose to leverage classifiers to measure the level of overlap between distributions.
\\
\noindent
\\
The relationship between classification performance and overlap is well studied in the literature, see \cite{garcia2007empirical,vuttipittayamongkol2021class}. In fact, it is well known that classification accuracy suffers from the overlap between classes.  The inverse problem of estimating the overlap using classification has been investigated as well. For example, the authors of \cite{johno2023decision} propose the use of decision trees to estimate the overlap between two probability distributions. In the context of our work, we only need an upper-bound for the overlap, since a low value of the upper-bound is a sufficient signal to discard a permutation. Therefore, we introduce a fast and scalable upper-bound measure of overlap which is the rate of error of a Quadratic discriminant analysis classifier. This approach is easily scalable as there is no need to train the classifier for the repeated use on different permutations. In particular, it suffices to calculate the empirical mean and covariance of the location vector of all players and we can deduce the parameters of the Quadratic discriminant analysis classifier to measure the overlap for any permutation. We also present and leverage a more complex classifier in the Bayesian Gaussian mixture classifier to provide tighter upper-bounds of the overlap.
\\
\noindent
\\
In addition, we propose an alternative approach to bypass the complexity arising from the cardinality of the permutation set. By selecting one player location per frame, we generate a new dataset where player locations are independent within each constructed frame. Each resulting player location follows a Gaussian mixture distribution, with all players sharing the same Gaussian mixture components. These common components represent the role distributions and the weights of the Gaussian mixture of each player represent the probability of the player being assigned to each role. By estimating the parameters of these shared component Gaussian mixtures, we are able to produce an initial estimate of the role locations and the average role assignment permutation. Furthermore, this estimation enables us to constrain the space of possible permutations, facilitating the estimation of the complete model.
\\
\noindent
\\
Our framework is highly valuable to practitioners at various levels, providing a systematic methodology to estimate team formations during games while accounting for potential position interchanges. For example, this enables the construction of detailed tactical profiles for each team by analyzing the formation clusters employed in their games. Furthermore, our model-based approach allows for an in-depth analysis of role assignment probabilities at a granular level. The inclusion of a regime variable also permits the automatic segmentation of games, enabling the detection of multiple formations and pinpointing the exact moments a change of structure occurs. Additionally, this framework is beneficial for scouting purposes, as the estimation of role assignments provides valuable insights into a player's abilities. Specifically, it helps identify players capable of assuming multiple roles during a game, thereby demonstrating their tactical flexibility and overall performance.
\\
\noindent
\\
The paper is organized as follows. In Section~\ref{section:model}, we present our approach to model the spatial distribution of players as well as the estimation procedure. We include the methodology used to select relevant permutations. In Section \ref{section:results}, we present applications of our methodology. First, we demonstrate how it is robust in the presence of overlap between role distributions. Then, we display an example of estimated formation in the game of Rennes, a randomly selected Ligue 1 team. We show how the model with multiple regimes can help extract  relevant insights on the game dynamics. Then, we include examples of analysis that can be derived using our framework such as the clustering of formations in the games of all Ligue 1 teams. In Section \ref{sec:overlaptoselect} of the Appendix, we present a detailed explanation of the overlap estimation as well as statistical guarantees. We include details on the Expectation-Maximization algorithm in the presence of permutations in Section~\ref{sec:em_algo}.

%% file: sections/methodology.tex
\section{Modeling tracking data}
\label{section:model}
 This section presents the framework we use to estimate football formations and identify player swaps. It includes a description of the estimation method based on the shared component Gaussian Mixture as well as the permutation selection procedure.
\subsection{Data}
\begin{figure}[t!]
    \centering
    \includegraphics[width=0.9\textwidth]{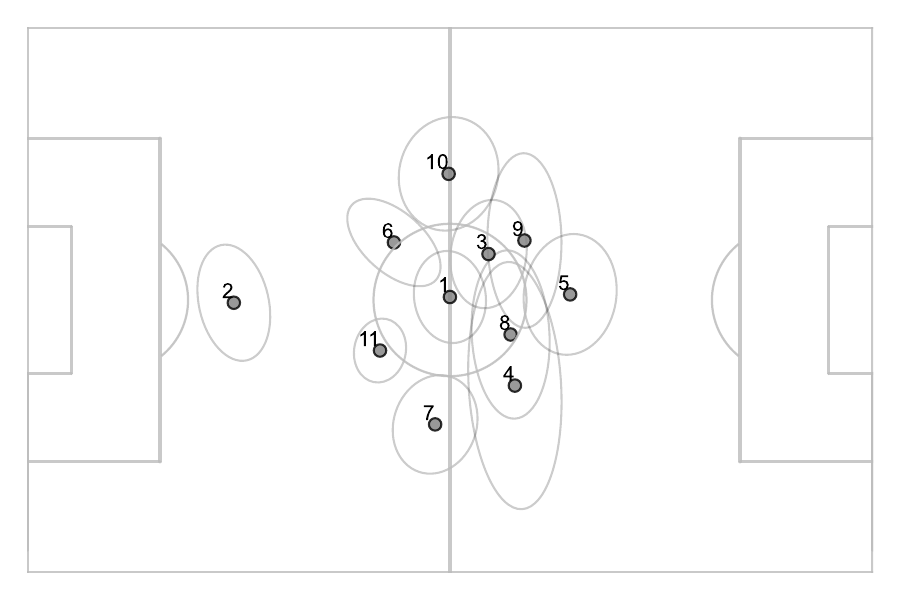}
    \vspace*{-.5em}
    \caption{Formation estimated using the naive approach with player locations after normalization. Each ellipse corresponds to the empirical covariance of a player's normalized position, 
    drawn at one standard deviation around the empirical mean. The locations are taken from the frames of Rennes in the game against Nantes in May 2022.}\label{fig:naivenormalized}
\end{figure}

The tracking dataset is provided by Stats Perform and lists $100$ games from the 2021-2022 Ligue 1 season with the trajectories of the players at a frequency of 25 frames per second. Each team has between 8 and 11 games in the dataset. Additionally, at each frame a value is provided specifying which team has possession of the ball, when the possession is assigned.
\begin{figure}[t!]
    \centering
    \includegraphics[width=0.9\textwidth]{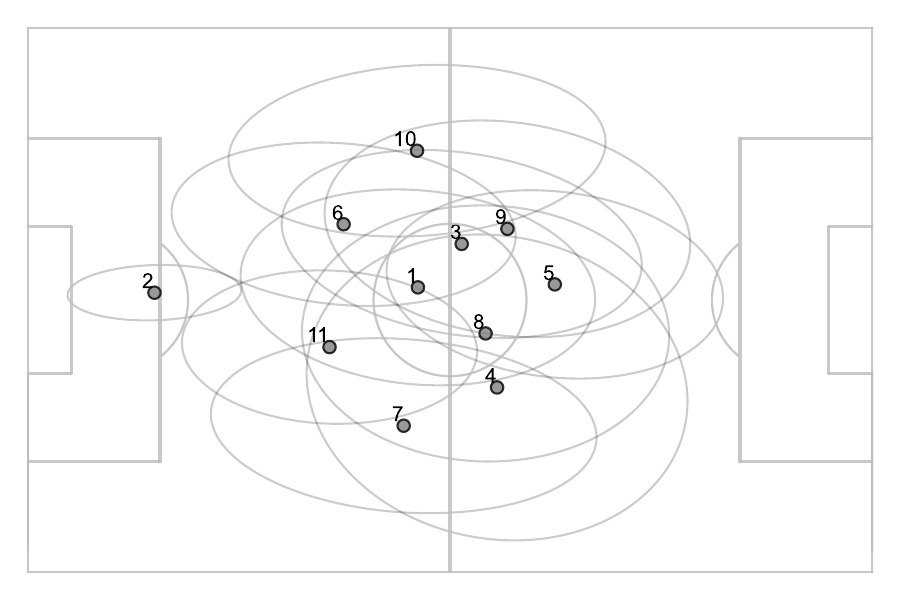}
    \vspace*{-.5em}
    \caption{Formation estimated using the naive approach with player locations without normalization. Each ellipse corresponds to the empirical covariance of a player position, 
    drawn at one standard deviation around the empirical mean. The locations are taken from the frames of Rennes in the game against Nantes in May 2022.}\label{fig:naive}
\end{figure}
\\
\noindent
\\
For each team and game, we build a list of location vectors $Y$ in $\mathbb{R}^{d\times 2}$ where $d=11$. This vector lists the positions of all players, with component $i$ corresponding to player of identity $i$ at a given frame. As the analysis on player-role assignment is sensitive to the identity of players, the components of the location vector have to be associated with the same player in all the frames. For this reason, we segment the match into sequences of frames where no substitutions has occurred and discard any segments shorter than five minutes to ensure sufficient data for reliable analysis. Furthermore, the tracking data is sub-sampled to reduce the time and space complexity, keeping one frame every 5 frames. This is also because consecutive frames at a $25$ frames per second frequency yield little additional information frame-to-frame. Moreover, the frame data is rotated when necessary with the convention that the team attacks the right side. 
\\
\noindent
\\
Figure~\ref{fig:naive} shows an example of the estimated formation using the naive approach, which is based on estimating the empirical mean and covariance matrix of each player. We observe that the player locations show a large amount of overlap. Part of this overlap is due to the fact that the team moves as a block and all players cover all areas of the pitch. Hence, to separate global pitch movement from role-based positioning, we normalize the locations in each frame. This is done by subtracting the mean location in the frame and dividing by the standard deviation along each axis. This allows us to focus on the shape and relative positions of players rather than their absolute locations. Figure~\ref{fig:naivenormalized} shows the formation estimated using normalized frames, but keeping a naive fixed assignment of roles to players. While this figure display a more structured formation, we can still identify overlapping player distributions. To go beyond these naive visualizations and formally capture the team’s underlying structure, we introduce a probabilistic model for role assignment. Our modeling effort aims to disentangle role swaps and regime changes from the player locations, and to extract  clear formation estimates as well as a measure of role assignment probability. 

\subsection{A model for role assignment}
\label{sec:modelperm}
\noindent
In this section, we introduce a setting to model the observed locations $Y$ of players at a given frame. This is a simpler setting with a single regime, to which we add another latent variable to incorporate multiple phases of play in Section~\ref{sec:regime}.
For $k=1,\dots,d$, $Y_k$ is the location vector in $\mathbb{R}^2$ of the player with identity $k$. We assume the existence of roles $k=1,\dots,d$, where each role is responsible for covering some area on the pitch centered around $\mu_k$ and of width characterized by a covariance matrix $\Sigma_k$. We assume further that at every frame, the roles are distributed to players randomly following some unknown distribution on permutations. The model can be written as follows:
\begin{equation}
\label{eq:model}
    Y=\Pi X,
\end{equation}
where  $X$ is the vector in $\mathbb{R}^{d\times 2}$ of role locations, and $\Pi $ is a random permutation matrix independent from $X$. The distribution of $\Pi$ is determined by the probability $w_Q$ of each permutation matrix $Q$. Additionally, $X_k$ is drawn independently from a normal distribution $\mathcal{N}(\mu_k,\Sigma_k)$ for $k=1,\dots,d$. In this work, $Y$ and $X$ are random variables in $\mathbb{R}^{d\times 2}$ that represent player and role locations respectively. While we keep them in their matricial representation for clarity in exposition, they are implicitly identified with their vectorized representations in $\mathbb{R}^{2d}$ in density calculations. The goal is to estimate simultaneously the law of $\Pi$ and the areas associated with every role, characterized by $(\mu_k,\Sigma_k)$. It should be noted that only the vector $Y$ is observed while $\Pi$ and $X$ are latent variables. 
\\
\noindent
\\
We recognize that this model is a special case of Gaussian mixtures, where the hidden component variable is in the form of a permutation matrix and where the component parameters are tied.  If we denote by $g(.,\mu,\Sigma)$ and $f(.,\mu,\Sigma,w)$ the density functions of $X$ and $Y$ respectively in $\mathbb{R}^{d\times 2}$, we see that $g$ is a multivariate normal distribution given by:
\begin{equation*}
    g(x,\mu,\Sigma)=\frac{1}{(2\pi)^d \sqrt{\prod\limits_{k=1}^d \det(\Sigma_k)}} \exp \left( -\frac{1}{2} \sum_{k=1}^d (\mu_k - x_k)^\top \Sigma_k^{-1} (\mu_k - x_k) \right),
\end{equation*}
 for $x$ in $\mathbb{R}^{d\times 2}$. As for $f$, it is a mixture of Gaussians given by: 
\begin{equation*}
    f(y,\mu,\Sigma,w)=\sum_{Q\in\mathcal{P}_d} w_Q g(Q^\top y,\mu,\Sigma),
\end{equation*}
for $y$ in $\mathbb{R}^{d\times 2}$. Here, \(\mathcal{P}_d\) denotes the set of all \(d \times d\) permutation matrices and $w=(w_Q)_{Q\in \mathcal{P}_d}$ is the vector of all permutation probabilities. Thus, the log-likelihood for independent observations $y^{(1)},y^{(2)},\dots,y^{(n)}$ is given by:

\begin{equation}
\label{eq:likelihood1}
\small \mathcal{L}_\theta(\mathbf{y}) = \sum_{i=1}^n \log \left[ \sum_{Q \in \mathcal{P}_d} w_Q \frac{1}{(2\pi)^d \sqrt{\prod\limits_{k=1}^d \det(\Sigma_k)}} \exp \left( -\frac{1}{2} \sum_{k=1}^d (\mu_k - (Q^\top y^{(i)})_k)^\top \Sigma_k^{-1} (\mu_k - (Q^\top y^{(i)})_k) \right) \right],
\end{equation}
\\
\noindent
where $\theta=\left ((w_Q)_{Q\in \mathcal{P}_d},(\mu_k)_{k\leq d},(\Sigma_k)_{k\leq d} \right )$ are the concatenated parameters. In this context, we assume that frames within the same game are independent, a simplification that facilitates the modeling process. 
\\
\noindent
\\
The maximization of the log-likelihood poses two challenges. Firstly, the presence of latent variables makes the likelihood function non-convex and difficult to optimize directly. Secondly, it involves the storage and estimation of $d!-1$ parameters which is computationally not feasible. Regarding the first challenge, we recognize that our model resembles a Gaussian Mixture Model (GMM) with interconnected component centers due to role assignments. Consequently, we employ the Expectation-Maximization (EM) algorithm to estimate the parameters, leveraging its ability to handle latent variables through alternating expectation and maximization steps, see Section~\ref{sec:em_algo} for more details. To address the second challenge, we introduce sparsity in the set of considered permutations, effectively reducing the parameter space by retaining only the permutations with significant probabilities.
\\
\noindent
\\
In addition to the assumption of the sparse permutation support, we also assume that there exists a primary role to player assignment. This is interpretable because every player is expected to occupy his designated role during a large portion of the game. Therefore, we aim at estimating the role parameters so that role $k$ is occupied by player $k$ with high probability. The parametrization in our model is identifiable up to a permutation of roles and subsequent adjustment of the distribution over permutations, see Section~\ref{sec:identifiability} of the Appendix for the proof. Among these equivalent solutions, we seek the one where the role indices and player identities are aligned. As a result, our assumption implies that the probability of the identity permutation \( P(\Pi = I_d) \) is non-negligible. We achieve this alignment of indices through two strategies: Initializing role $k$ parameters close to the locations of player $k$, and selecting the likely permutations with the assumption that $P(\Pi=I_d)$ is large. 
\subsection{Shared component GMM}
\label{sec:estimation_iid}
One approach to address the cardinality of the set of permutations is to eliminate dependencies among players by randomly selecting one player per frame. We generate a new dataset, in which each frame consists of player locations drawn from different frames, ensuring that player components are independent. By doing so, we relax some of the structural constraints imposed by permutation-based modeling, allowing for example two players within a constructed frame to occupy the same role.
\\
\noindent
\\
Formally, consider a set of independent realizations \( Y^{(i)} \) for \( i = 1, \dots, n \), where \( Y^{(i)} \) lists the player locations in frame $i$. To ensure that each player is selected an equal number of times, we use only the first
\(
11 \left\lfloor \frac{n}{11} \right\rfloor
\)
 frames. Then, at each frame, we randomly select one player location, with the constraint that all players are selected the same number of times. For each player \( l = 1, \dots, 11 \) and for \( j = 1, \dots, \left\lfloor \frac{n}{11} \right\rfloor \), let \( i_{jl} \) denote the frame index where player \( l \) is selected the $j^{th}$ time. The sets \( \{i_{jl},\ j=1,\dots, \left\lfloor \frac{n}{11} \right\rfloor\} \) for $l=1,\dots, 11$ form a random partition of the data into equally sized subsets for each player. For each index \( j = 1, \dots, \left\lfloor \frac{n}{11} \right\rfloor \), we construct a new frame \( Z^{(j)} \) with these locations
\begin{equation*}
    Z^{(j)}=\left (Y^{i_{jl}}_{l}\right )_{l=1,\dots,11}.
\end{equation*}
In this setup, the components of \( Z^{(j)} \) are independent because they come from distinct and independent frames. Moreover, the constructed frames are also independent. This approach maintains information about the role parameters \( \mu_k \) and \( \Sigma_k \) while removing the dependency structure introduced by the permutation matrix \( \Pi \). For each component \( l = 1, \dots, 11 \), the random variable \( Z_l^{(j)} \) has the density function
\begin{equation*}
    f_l(z) = \sum_{k=1}^{11} \pi_{l,k}\ \mathcal{N}(z \mid \mu_k, \Sigma_k),
\end{equation*}
where \( \pi_{l,k} = \mathbb{E}(\Pi_{l,k}) = P(\Pi_{l,k} = 1) \) is the probability that role \( k \) is assigned to player \( l \). Thus, each component of \( Z \) is an independent Gaussian mixture with shared component parameters across players. We estimate the weights and the shared components using the Maximum-likelihood estimation. Through independence, the log-likelihood can be written as follows:
\begin{equation}
\label{eq:likelihoodindependent}
\small \mathcal{L}_\theta(\mathbf{z}) = \sum_{j=1}^ {\left \lfloor \frac{n}{11} \right \rfloor} \sum_{l=1}^{11}\log \left[ \sum_{k=1}^{11} \pi_{l,k} \frac{1}{2\pi \sqrt{\det(\Sigma_k)}} \exp \left( -\frac{1}{2} (z^{(j)}_l-\mu_k)^\top \Sigma_k^{-1} (z^{(j)}_l-\mu_k) \right) \right].
\end{equation}
Fitting the shared component mixture model using the EM algorithm provides an initial estimate for the role centers \( \mu_k \) and covariance matrices \( \Sigma_k \). It also gives an estimate of the average permutation probabilities \( \pi_{l,k} \).
\\
\noindent
\\
However, this method ignores some important aspects of the original model, specifically the structure at the frame level where there is a one-to-one correspondence between players and roles. This is because we only retain one player location from each original frame. As a result, the estimated average permutation matrix \( \pi \) will have rows that sum to 1, reflecting that each player is assigned one role at each frame, but there are no constraints on the columns of \( \pi \). This estimation procedure does not incorporate the fact that each role is assigned to exactly one player per frame. Furthermore, one of the additional drawbacks is that we discard a large portion of the data. To reflect the true constraint that each player is assigned to exactly one role and estimate the probability of each permutation, we must use the complete model with permutations.
\\
\noindent
\\
The parameters retrieved with this model are useful in our procedure at two distinct levels. First, they provide an initial estimate of the formation of the team. This estimation can be used as initialization for the Gaussian mixture with permutations to speed up convergence. Second, the initial estimate of the expected permutation matrix is used to select the relevant permutations. In fact, given a permutation matrix $Q$, its probability is inferior to $\pi_{l,k}$ for all indices $l,k$ such that $Q_{l,k}=1$.
\subsection{Selecting relevant permutations}
\label{sec:selectingpermutation}   
Reducing the cardinality of the permutation set is crucial for computational feasibility and modeling efficiency, especially in high-dimensional applications like football tracking data.  In fact, the computation for the optimization in Equation \eqref{eq:likelihood} becomes computationally infeasible if we include the entire set of permutations, whose cardinality is $d!$. In the context of football tracking data, it is reasonable to assume that the support of the distribution of permutations is parsimonious. This motivates the introduction of a procedure to select relevant permutations that exhibit statistically significant evidence of having sufficient probability. We propose the combined use of two criteria:
\paragraph{Overlap criterion}
The underlying idea is based on the fact that if a permutation matrix $Q$ has a sufficient probability, then the densities of the random variables $Y$ and $QY$ exhibit substantial overlap. In the context of our model, the following proposition links the overlap and the probability of a permutation. 

\begin{proposition}
\label{lemma:overlap_permutation}
Let $Q$ be a permutation matrix and consider the random variables $Y$ and $QY$ with respective densities with respect to the Lebesgue measure $p$ and $q$. Then \begin{align*}
p(y)&\geq P(\Pi=I_d) f_X(y) + P(\Pi=Q) f_{QX}(y) ,
\end{align*}
and 
\begin{align*}
q(y)&\geq P(\Pi=Q^\top) f_X(y) + P(\Pi=I_d) f_{QX}(s),
\end{align*}
where $f_X$ and $f_{QX}$ are the density functions of $X$ and $QX$ respectively.
In particular, the overlap between $Y$ and $QY$ verifies
\begin{equation}
\label{eq:overlap_Q}
    v(p,q)\geq \min (P(\Pi=Q^\top),P(\Pi=I_d))+\min (P(\Pi=Q),P(\Pi=I_d)),
\end{equation}
where $v(p,q)$ is the overlap between $p$ and $q$ defined as $$v(p,q)=\int_{R^{d\times 2}} \min (p(x),q(x))dx.$$
\end{proposition}
\noindent
Here, the density functions in $\mathbb{R}^{d\times2}$ are identified with the ones for the flattened vectors in $\mathbb{R}^{2d}$. We observe that the role-player permutation can be a source of overlap between $Y$ and $QY$. By hypothesis, we want to select only permutations with a probability above some threshold level. To do so, we fix an overlap threshold level $o^{tresh}$, and only select permutations $Q$ such that $Y$ and $QY$ display an overlap above $o^{tresh}$. Furthermore, we make the hypothesis that the probability of the identity permutation  is non-negligible and satisfies $P(\Pi=I_d)>o^{tresh}$. This is a natural hypothesis as players are assumed to occupy the location related to their role a minimal amount of time. The selection criterion is simplified using Equation~\eqref{eq:overlap_Q}
\begin{equation*}
   v(p,q)\leq o^{tresh} \Rightarrow  P(\Pi=Q)+P(\Pi=Q^\top) \leq o^{tresh}.
\end{equation*}
Therefore, the threshold on overlap $o^{tresh}$ that we fixed can be interpreted as a lower bound on the probability of the permutation and the probability of its inverse. In the special case where $Q$ is a transposition or a group of disjoint transpositions, we have $Q=Q^\top$ and the necessary condition to discard a permutation becomes $ P(\Pi=Q)\leq \frac{o^{tresh}}{2}$.
\begin{remark}
\label{remark:overlap}
    The presence of overlap is a necessary but not sufficient condition for a permutation to have a non-negligible probability. In particular, for some permutation matrix $Q$ there can be overlap between $QY$ and $Y$ for different reasons, given the complex interactions between players and roles.
\end{remark}
\noindent
Among the reasons mentioned in Remark~\ref{remark:overlap}, there is the intrinsic overlap between all normal distributions. This overlap is difficult to estimate since the roles and their parameters are not known beforehand. If two roles are associated to Gaussians that are located close to one another, the locations of the players associated to them naturally show some overlap and the transposition between them is automatically selected. Another source of overlap can be due to other permutations. In particular, if $Q$ and $\Tilde{Q}$ are permutation matrices such that $\Tilde{Q}$ and $Q\Tilde{Q}$ have a non null probability, $Y$ and $QY$ still show overlap despite the permutation matrix $Q$ having probability zero. This is because $Y$ conditionally to $\Pi=Q\Tilde{Q}$ is equal to $Q\Tilde{Q}X$, and similarly $QY$ conditionally to $\Pi=\Tilde{Q}$ is equal to $Q\Tilde{Q}X$. 
\paragraph{Using an initial estimate of $\mathbb{E}(\Pi)$}
Section~\ref{sec:estimation_iid} provides a procedure to get an initial estimate of $\mu_k,\Sigma_k$ as well as the expectation of $\mathbb{E}(\Pi)$. This estimate can be leveraged to determine the possible assignments of roles to players as those associated with a positive entry in the estimated matrix $(\hat{\pi}_{l,k})_{l,k\leq 11}$. This procedure can be used prior to the overlap criterion to significantly reduce the space of relevant permutations. In particular, we fix a probability threshold $p^{tresh}$ and only select permutation matrices $Q$ such that $\hat{\pi}_{l,k}\geq p^{tresh}$ for all $l,k$ such that $Q_{l,k}=1$.
\\
\noindent
\\
We employ a combination of both methods to efficiently select permutations prior to the EM-algorithm estimation. As a first step, the estimated $(\hat{\pi}_{l,k})_{l,k\leq 11}$ using the shared component GMM provides a list of relevant permutations. As a second step, we discard those that show a small amount of overlap. Details on how the overlap is estimated can be found in Section~\ref{sec:overlaptoselect} in the Appendix.

\subsection{Introducing a regime latent variable}
\label{sec:regime}
The permutation selection algorithm described in the previous section allows the efficient use of Maximum-likelihood estimation on the Gaussian mixture with permutations to determine a team's formation as well as the rate of player swaps. An additional advantage of this model is that it can be extended to incorporate regime changes within a game. 
\\
\noindent
\\
A game of football is subject to permanent or temporary changes in formation. To account for such variations and avoid restricting the shape of the team to one single formation for an entire game, we generalize the model to incorporate regime switches. Specifically, we consider the joint distribution of the permutation matrix \(\Pi\) and the role-centered locations \(X\) as a mixture of distributions, each corresponding to a distinct tactical regime. To formalize this, we introduce an additional discrete random variable \(R\) representing the regime, such that for each \(r = 1,\dots,l\):

\[
    X_k \mid R = r \sim \mathcal{N}(\mu_{r,k}, \Sigma_{r,k}),
\]
and the permutation matrix \(\Pi\) is conditioned on the regime \(R = r\), with
\[
    P(\Pi = Q \mid R = r) = w_{r,Q}.
\]
\\
\noindent
Here, each regime \(r\) represents a specific formation characterized by its own set of mean locations \(\mu_{r,k}\) and covariance matrices \(\Sigma_{r,k}\). The permutation probabilities \(w_{r,Q}\) capture the likelihood of assigning players to roles within each regime. By introducing the regime variable \(R\), the model can flexibly adapt to different tactical phases of the game, such as transitioning from an offensive to a defensive formation, thereby providing a more comprehensive and dynamic representation of team strategies. In this setting, the log-likelihood can be written as:
\begin{equation}
\label{eq:likelihood}
\begin{aligned}
\mathcal{L}_\theta(\mathbf{y}) = \sum_{i=1}^n \log \Bigg[ & \sum_{r=1}^l v_r \sum_{Q \in \mathcal{P}_d} w_{r,Q} \frac{1}{(2\pi)^d \prod_{k=1}^d \sqrt{\det(\Sigma_{r,k})}} \\
& \times \exp \left( -\frac{1}{2} \sum_{k=1}^d (\mu_{r,k} - (Q^\top y^{(i)})_k)^\top \Sigma_{r,k}^{-1} (\mu_{r,k} - (Q^\top y^{(i)})_k) \right) \Bigg] ,
\end{aligned}
\end{equation}
where $v_r=P(R=r)$ is the probability of the regime $R=r$.
\\
\noindent
\\
The estimation of the parameters using independent samples is performed using the Expectation-Maximization (EM) algorithm. Detailed descriptions of the E-step and M-step within this context are provided in Section \ref{sec:em_algo} of the Appendix. Through this iterative procedure, we recover a set of parameters that offer direct insights into a team's style of play:
\begin{itemize}
    \item $\hat{v}_r$: The probability of each regime occurring during the game. This can also be seen as the expected proportion of frames where the regime occurs. 
    \item $\hat{w}_{r,Q}$: The probability of each permutation of roles occurring conditionally on the regime $r$. 
    \item \(\hat{\mu}_{r,k}\) and \(\hat{\Sigma}_{r,k}\): The estimated mean position and covariance matrix for each role \(k\) within regime \(r\), respectively.
\end{itemize} 
\noindent
Using these estimates, we can derive model-based metrics to evaluate the game in terms of regime formation and player switching probability:
\begin{itemize}
    \item \textbf{Expected assignment of roles to players in each regime}: This quantity can be found in the estimated expectation of the hidden permutations conditionally to each regime state. 
    \begin{equation}
    \label{eq:avg_permutation}
        \hat{\pi}_r=\sum_{Q\in \mathcal{P}_d} \hat{w}_{r,Q} Q .
    \end{equation}
     Note that the sum is performed over the selected support of the permutation distribution. The entry \((l,k)\) of the resulting matrix \(\hat{\pi}_r\) represents the probability that player \(l\) occupies role \(k\) under regime \(r\).
    \item \textbf{Frame regime probability}: For each frame \(i\) and regime \(r\), the probability that the game is in regime \(r\) is computed as: 
    \begin{equation}
    \label{eq:frame_regime_probability}
        \hat{v}_{i,r}=\frac{\sum\limits_{Q\in \mathcal{P}_d} g\bigl(Q^\top y^{(i)};\hat{\mu}_r, \hat{\Sigma}_r \bigr) \hat{w}_{r,Q}\hat{v}_r}{\sum\limits_{r=1}^l\sum\limits_{Q\in \mathcal{P}_d}g\bigl(Q^\top y^{(i)};\hat{\mu}_r, \hat{\Sigma}_r \bigr) \hat{w}_{r,Q}\hat{v}_r},
    \end{equation}
    where \( g(.; \mu, \Sigma) \) denotes the density function of $X$ introduced in Section \ref{section:model}. This expression calculates the posterior probability of regime \(r\) conditionally on the observed location in frame \(i\).

    \item  \textbf{Frame assignment of roles probability}: For each frame $i$ and regime $r$, the probability of permutation $Q$ is given by:
    \begin{equation}
    \label{eq:frame_permutation_probability}
        \hat{w}_{i,r,Q}=\frac{g\bigl(Q^\top y^{(i)};\hat{\mu}_r, \hat{\Sigma}_r \bigr) \hat{w}_{r,Q}}{\sum\limits_{Q\in \mathcal{P}_d} g\bigl(Q^\top y^{(i)};\hat{\mu}_r, \hat{\Sigma}_r \bigr) \hat{w}_{r,Q}}.
    \end{equation}
    In particular, this permits the estimation at the frame level of the probability of no swap happening, providing a measure of rigidity in placement of the team.
\end{itemize} 

%% file: sections/results.tex
\section{Numerical results and discussion}
\label{section:results}
In this section, we display some examples of estimations using the 1-regime and multiple regime models and present some results that can be derived using this modeling. We start with a simulation study to underline the performance of the estimation procedure in the presence of inherent overlap between roles.
\\
\noindent
\\
Then, we display estimation results on the tracking data of Stade Rennais, a randomly selected Ligue 1 team, in its game against Nantes. The estimation procedure can be described as follows: 

\begin{enumerate}
    \item We fit the shared component GMM using the EM algorithm and retrieve estimates $\hat{\mu}^{(shared)}$,$\hat{\Sigma}^{(shared)}$ of the roles parameters as well as the average permutation $\hat{\pi}^{(shared)}$.
    \item We construct a list of permutations $Q$ 
 that satisfy  $\hat{\pi}^{(shared)}_{l,k}\geq p^{tresh}$ for all $l,k$ where $Q_{l,k}=1$, with $p^{tresh}=2.5\%$. 
 \item We discard from these permutations those that show a level of overlap less than $o^{tresh}=5\%$. The estimation of the asymptotic overlap upper-bound is done using a classifier following Section~\ref{sec:overlaptoselect} in the Appendix, with a confidence level of $1-\alpha=95\%$. This is done in two steps: first, we use the faster QDA classifier explained in Section~\ref{sec:QDA}. Then, we perform a second layer of selection by using a tighter upper-bound with the Bayesian Gaussian Mixture classifier, see Section~\ref{sec:bayesianmixture} in the Appendix. The Gaussian mixture is trained on part of the frames that are removed during sub-sampling.
 \item Estimate the GMM with permutations using the selected permutations.
\end{enumerate}
\begin{remark}
 The choice of $o^{tresh}=5\%$ strikes a balance between being low enough to discard the negligible permutations and large enough to not select a large number of permutations. It is given a larger value than $p^{tresh}=2.5\%$ for two reasons. First, a permutation discarded with the overlap criterion has a probability at most $o^{tresh}=5\%$ in the worst case. In the case where the permutation is a transposition or product of disjoint transpositions, the maximum probability is lower and given by $\frac{o^{tresh}}{2}=2.5\%=p^{tresh}$. The second reason is due to the conservative nature of the overlap criterion. In particular, overlap can arise due to different reasons to the permutation probability as stated in Remark \ref{remark:overlap}.
\end{remark}

\paragraph{Parameter initialization} We emphasize the importance of a sensible initialization for the EM algorithm. Due to the model’s invariance by permutation, swapping two roles in the parameter set while accordingly adjusting the distribution on permutations does not change the distribution of \(Y\). Therefore, we initialize the model to encourage the convergence to a solution where role $k$ is assigned with high probability to player $k$. To achieve this, we choose the following initialization strategy for each model:
\begin{enumerate}
    \item  \textbf{Shared component GMM}: The Gaussian component $k$ is initialized with the empirical average and covariance matrix of player $k$. Additionally, we initialize the weights $\pi_{l,k}$ as $\pi_{l,l}=0.5$ and $\pi_{l,k}=0.05$ for $k\neq l$. As a result, the player $k$ initially has a large weight on role $k$ compared to the other roles.
    \item \textbf{One-regime model with permutations}: The role means and covariance matrices are initialized using the estimates derived from the shared component model. Additionally, the permutation weights are initialized based on the maximum probabilities they can have according to the entries of the estimated average permutation $\hat{\pi}^{(shared)}$. Specifically, for each permutation matrix \( Q \), its initial weight is proportional to the minimum of \( \hat{\pi}^{(\text{shared})}_{l,k} \) for all indices \( l, k \) where \( Q_{l,k} = 1 \). Formally, this can be expressed as:
\[
w_{Q} \sim \min \left\{ \hat{\pi}^{(shared)}_{l,k} \mid Q_{l,k} = 1 \right\}.
\]
\end{enumerate}

\subsection{Test on simulated data}
\begin{figure}[t!]
    \centering
    \begin{subfigure}[b]{0.45\textwidth}
        \centering
        \includegraphics[width=\textwidth]{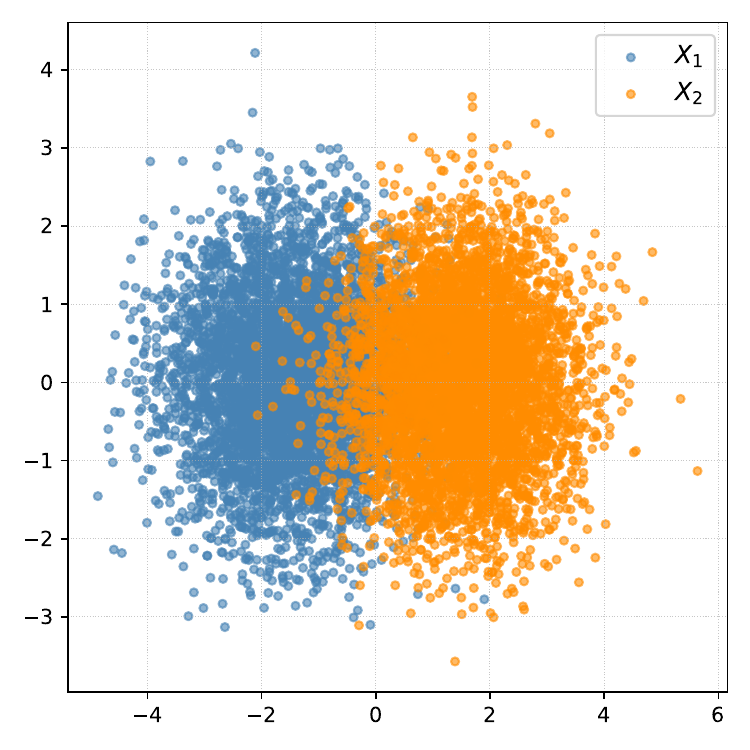}
        \caption{Scatter plot of the samples of $X$.}
        \label{fig:original_data}
    \end{subfigure}
    \hfill
    \begin{subfigure}[b]{0.45\textwidth}
        \centering
        \includegraphics[width=\textwidth]{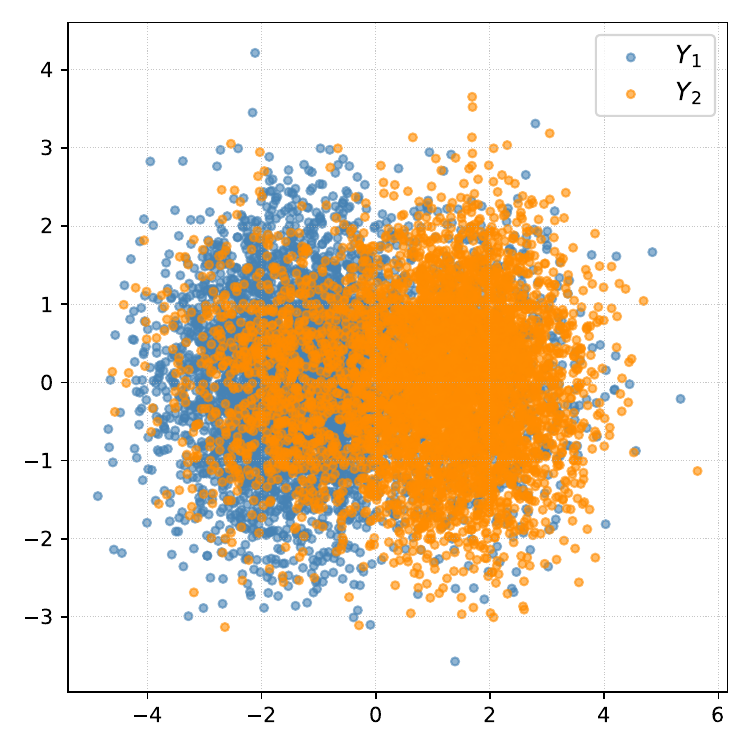}
        \caption{Scatter plot of the samples of $Y$.}
        \label{fig:swapped_data}
    \end{subfigure}
    \caption{Scatter plots of generated samples with mean vectors \( \mu_1 = -\mu_2 = (1.5, 0) \). The left panel displays the original distributions of \( X \), while the right panel illustrates the distributions of \( Y \) after applying a permutation with 20\% probability.}
    \label{fig:gaussian_comparison}
\end{figure}

\begin{figure}[t!]
    \centering
    \includegraphics[width=0.8\textwidth]{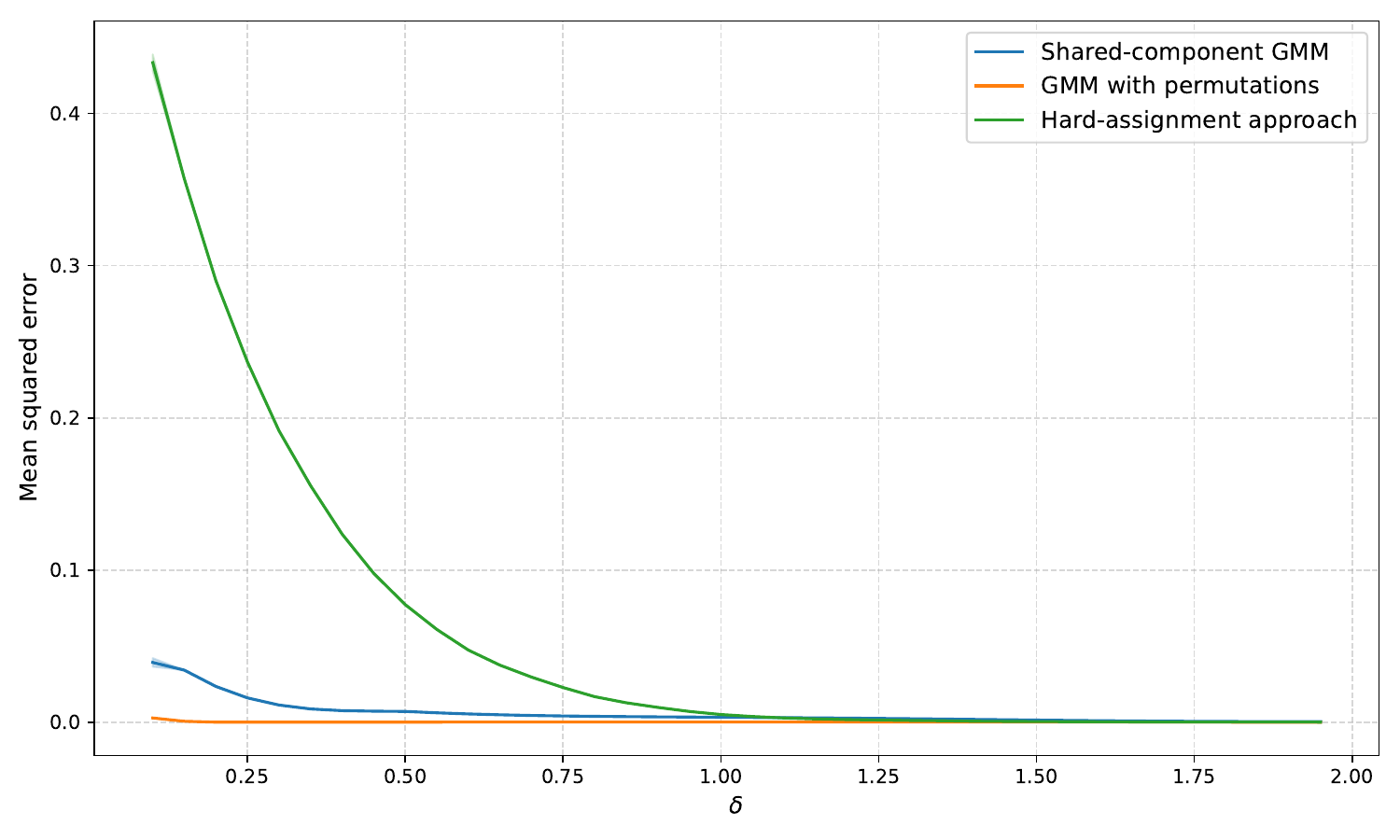}
    \caption{Evolution of the mean squared error in the estimation of $(\mu_1,\mu_2)$ of the three methods across varying values of $\delta$.}
    \label{fig:error_comparison_mu}
\end{figure}

The goal of our model is to explain the overlap between player location using the hidden permutations and regime variables. However, the overlap can also be inherently present between roles, where two roles are supposed to cover a common area. It is therefore important to determine the robustness of our estimation procedure to the presence of overlap between roles. 
\\
\noindent
\\
To achieve this, we perform an experiment on simulated data in a simple setting of two roles. Specifically, we generate samples from two Gaussian components in \(\mathbb{R}^2\), where:
\[
X_1 \sim \mathcal{N}(\mu_1, I_2)
\quad\text{and}\quad
X_2 \sim \mathcal{N}(\mu_2, I_2).
\]
with \(\mu_2 = -\mu_1=\left (\delta,0\right )\) . The choice of \( \delta=\frac{\|\mu_1 - \mu_2\|}{2}\) determines the level of overlap between the Gaussian components. The closer it is to zero, the closer the two Gaussian components are. We then apply a random permutation of the two components with probability \(p = 0.2\), yielding the observed player locations \(Y\). Figure~\ref{fig:original_data} displays an example of samples of $X_1$ and $X_2$ for \(\mu_1 = -\mu_2  = (1.5, 0)\), while Figure~\ref{fig:swapped_data} shows the observed locations $Y_1$ and $Y_2$ after the permutation. The presence of intrinsic overlap between roles poses a challenge to the estimation task where we look to disentangle the two components.
\\\noindent
\\
In this setting, we compare the accuracy of three estimators in approximating the true parameters $(\mu_1,\mu_2)$:
\begin{itemize}
    \item \textbf{The one-regime model with permutations}, as presented in Section~\ref{sec:modelperm}.
    \item  \textbf{The shared component GMM}, as introduced in Section~\ref{sec:estimation_iid}.
    \item \textbf{Hard-assignment approach}, following \cite{bialkowski2016discovering}. We initialize the estimates for each role as the empirical mean and covariance of each component of $Y$. Then iteratively, we solve the optimal assignment problem between players and roles for each data-point to maximize the log-likelihood and update the empirical mean and covariance of roles with the assigned points. 
\end{itemize} 
For each value of $\delta$ between $0.1$ and $2$, we draw \(n=5000\) i.i.d.\ samples of \(Y = (Y_1, Y_2)\) and perform the estimation of the roles parameters $\mu_1$ and $\mu_2$ using the three methods. The experiment is repeated $100$ times for every value of $\delta$ to compute the mean squared error in Figure~\ref{fig:error_comparison_mu}. Unsurprisingly, the hard-assignment approach is particularly sensitive to overlapping classes: when the roles are closer, it has difficulty assigning the data points consistently, thereby introducing bias. On the other hand, both the complete mixture model and the shared component approach are robust to the presence of intrinsic overlap. The slight drop in the shared component GMM’s accuracy is attributable to its sub-sampling scheme, which implies information loss. It should be noted that the hard-assignment method performs well with the absence of overlap. This is in line with the general fact that the K-means algorithm is equivalent to the EM algorithm in the limit where clusters are well separated, see \cite{kearns1998information}.  

\subsection{Visualizing team formation}
In this section, we illustrate the proposed method on the tracking data of Rennes, a randomly selected Ligue 1 team, in its game against Nantes on 11 May 2022. We focus on the frames from kickoff until the first substitution, sub-sampled by keeping one frame every five frames. Frames with missing players are excluded to ensure complete coverage of all eleven players. 

\subsubsection{One-regime estimation}
\begin{figure}[t]
    \centering
    \includegraphics[width=0.7\linewidth]{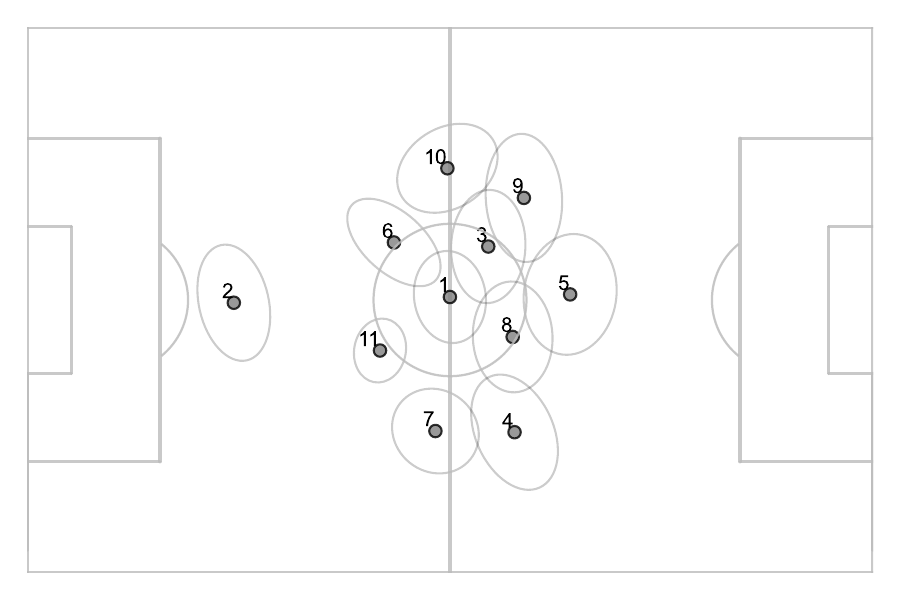}
    \caption{Formation estimated using the 1-regime model. Each ellipse corresponds to the empirical covariance of a role position, 
    drawn at one standard deviation around the empirical mean.}
    \label{fig:regular_em}
\end{figure}

\begin{figure}[htbp]
    \centering
    \begin{minipage}{0.48\linewidth}
        \centering
        \includegraphics[width=\linewidth]{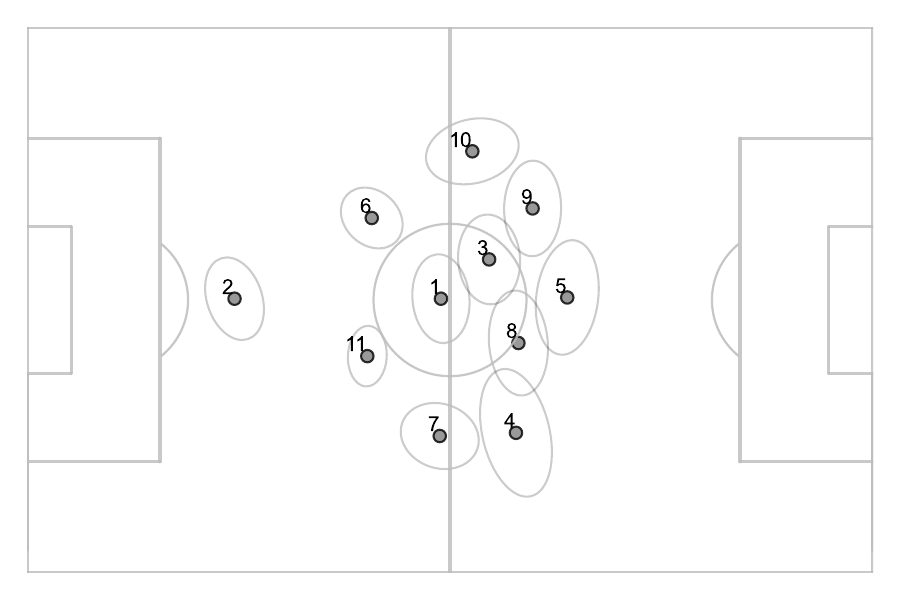}
        \caption*{(a) Formation in possession}
    \end{minipage}\hfill
    \begin{minipage}{0.48\linewidth}
        \centering
        \includegraphics[width=\linewidth]{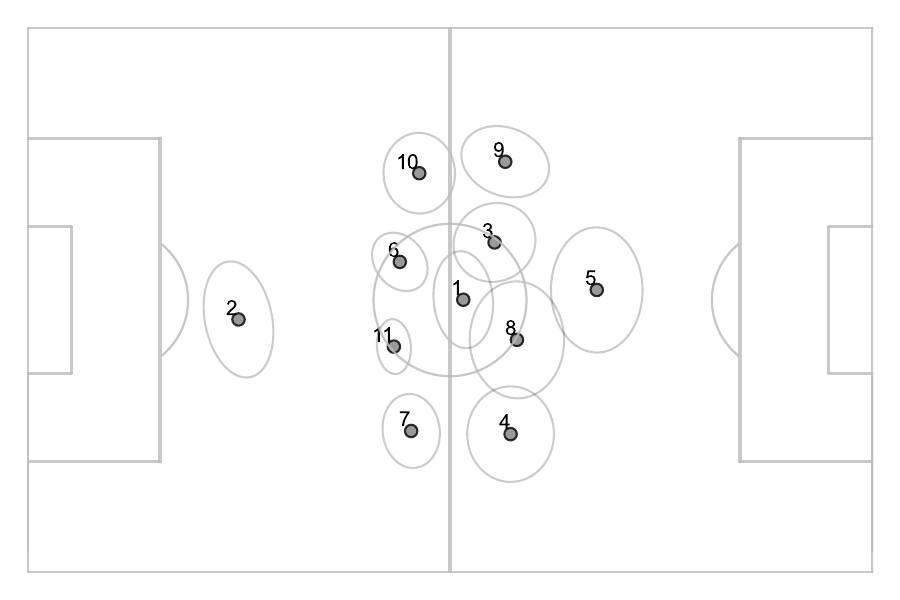}
        \caption*{(b) Formation out of possession}
    \end{minipage}
    \caption{Formations estimated using the 1-regime model in possession and out of possession. Each ellipse corresponds to the empirical covariance of a role position, 
    drawn at one standard deviation around the empirical mean.}
    \label{fig:em_attack_defense}
\end{figure}

Figure~\ref{fig:regular_em} displays the estimated formation with the mean location of each role and its covariance matrix under the 1-regime Gaussian mixture with permutations. Not surprisingly, Stade Rennais adopts a 4--3--3 shape, aligning with the official formation for this game. More importantly, we can see that the level of overlap between roles is substantially reduced compared to the naive approach in Figure~\ref{fig:naivenormalized}.
\\
\noindent
\\
A more refined approach is to separate frames according to ball possession. Here, the dataset is split into two subsets, one where the team under consideration holds possession (the attack subset) and another where the opponent controls the ball (the defense subset). Essentially, this is the same as manually determining regimes prior to estimating the formations. We apply our methodology to these subsets separately, and display the resulting formations in Figure~\ref{fig:em_attack_defense}. We can see that the fundamental 4--3--3 shape remains consistent across both phases, but we note subtle positional differences. For example, full-backs align more strictly with the center-backs when defending, whereas they advance wide and higher up the pitch when attacking. This is not surprising as out of possession, the line of defense has to be held strictly to enforce the off-line trap.

\subsubsection{Model with multiple regimes}
\begin{figure}[htbp]
    \centering
    \begin{minipage}{0.48\linewidth}
        \centering
        \includegraphics[width=\linewidth]{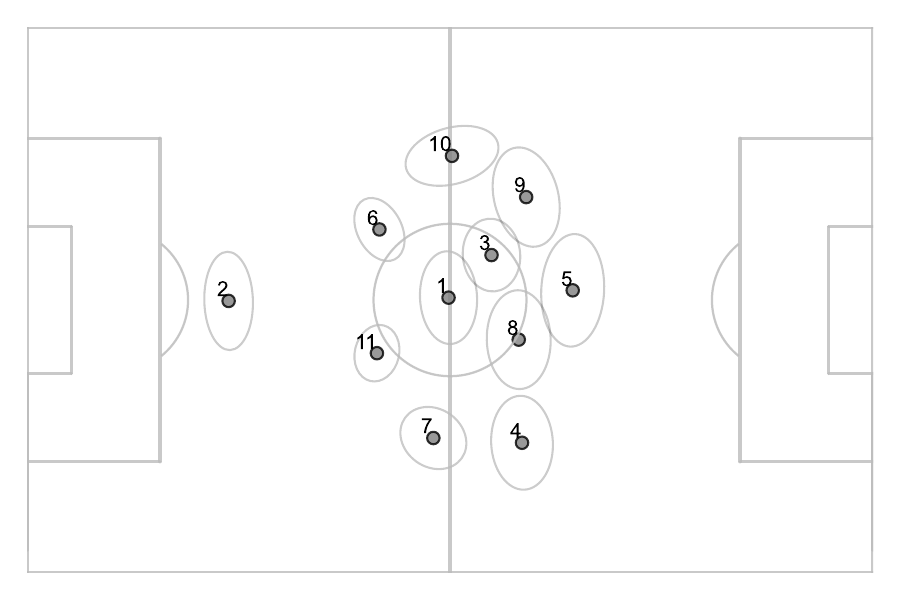}
        \caption*{(a) Regime 1 with probability 77.78\%.}
    \end{minipage}\hfill
    \begin{minipage}{0.48\linewidth}
        \centering
        \includegraphics[width=\linewidth]{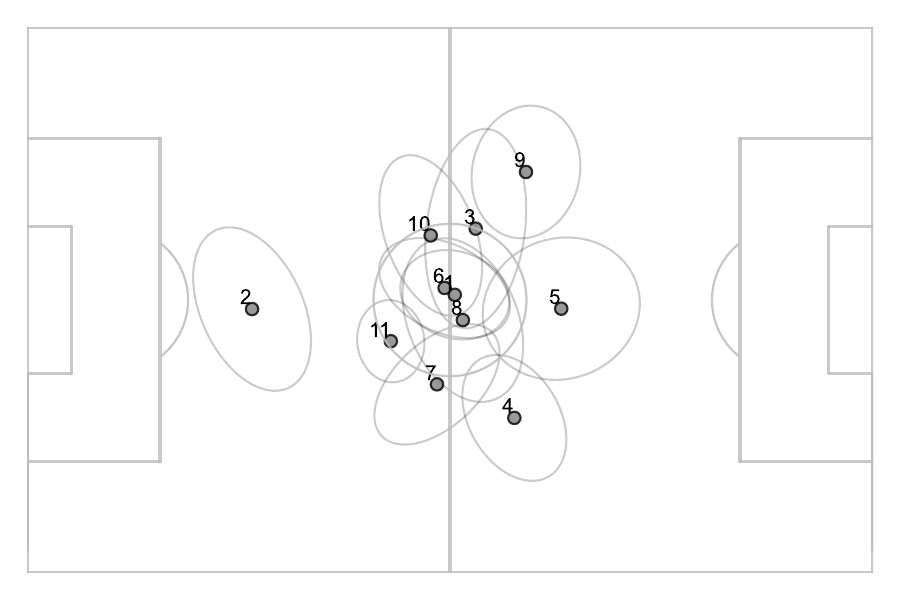}
        \caption*{(b) Regime 2 with probability 22.22\%.}
    \end{minipage}
    \caption{Formations estimated using the 2-regime model. Each ellipse corresponds to the empirical covariance of a role position, 
    drawn at one standard deviation around the empirical mean.}
    \label{fig:two_regime_em}
\end{figure}

Moving beyond a simple attack/defense dichotomy, we consider fitting a model with multiple regimes to determine distinct phases of play in an unsupervised way. Figure~\ref{fig:two_regime_em} displays the estimated formation in the two-regime model. In this approach, we initialize the formations of each regime using the ones estimated using the in-possession and out-of-possession fits. Surprisingly, the regime formations diverge from these initial estimates and yield a second regime with a very overlapping formation.  We hypothesize that this regime acts as an outlier cluster, grouping all frames where the team does not follow any clear structure. This is not surprising as some works reserve additional mixture components to capture outliers, see \cite{banfield1993model}. As a result, the formation retrieved in the first regime shows less overlap between roles than the one found in the case with only one regime.
\\
\noindent
\\
The second regime acting as an outlier cluster is consistent across the game segments we analyzed. To verify this, we measure the level of overlap between roles in the estimated formation using one and two regimes. Here, we choose the Bhattacharyya coefficient to measure the overlap between roles, see \cite{bhattacharyya1943measure}. It is defined between two probability densities $f_1$ and $f_2$ as:
\[
\mathrm{BC}(f_1, f_2) \;=\; \int_{\mathbb{R}^2} \sqrt{\,f_1(x)\,f_2(x)\,}\,dx.
\]
Given an estimated formation \((\mu, \Sigma)\), we compute the Bhattacharyya coefficient between \(\mathcal{N}(\mu_k, \Sigma_k)\) and \(\mathcal{N}(\mu_l, \Sigma_l)\) for every pair of distinct roles \(k, l\). We then use the average value of these coefficients as an indicator of overlap within the formation. A low Bhattacharyya coefficient suggests that the roles are well-separated and occupy distinct areas of the pitch. In contrast, a high coefficient indicates significant overlap between estimated roles, which may result either from truly intersecting roles or from the estimation process failing to disentangle role swaps from player locations.
\begin{remark}
   For this analysis, we use the Bhattacharyya coefficient as a measure of overlap instead of \( v(f_1, f_2) \), which is used in the permutation selection step. This choice is motivated by the fact that the Bhattacharyya coefficient admits a closed-form expression when the densities are Gaussian.
\end{remark}
\begin{figure}
        \centering
        \centering
            \includegraphics[width=0.6\linewidth]{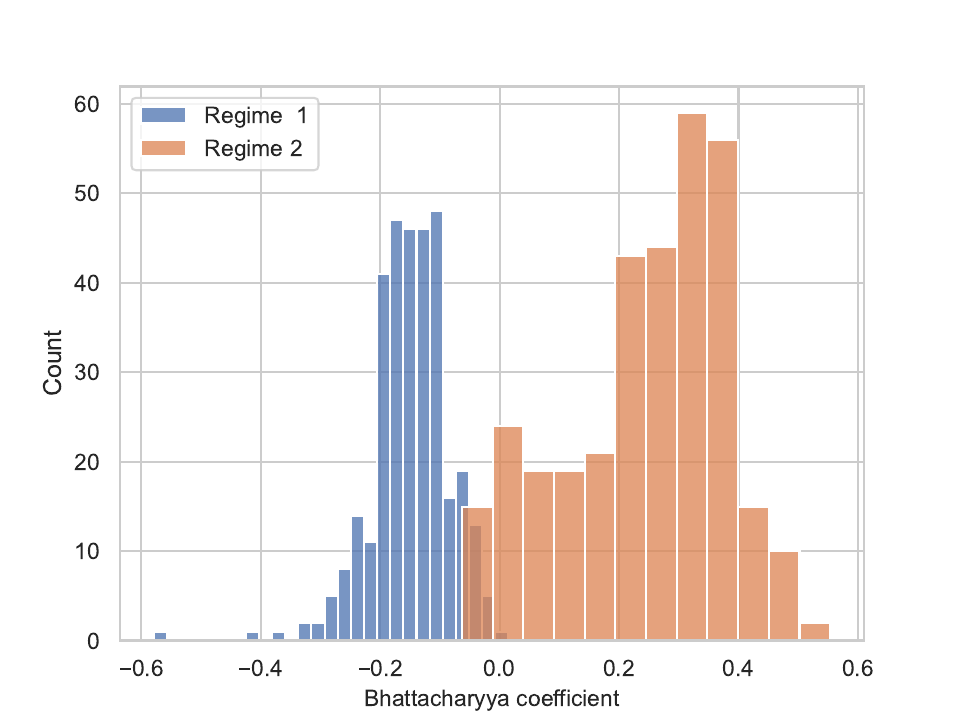}
    \caption{Distribution of $\mathrm{BC}_{2,1} - \mathrm{BC}_1$ and $\mathrm{BC}_{2,2} - \mathrm{BC}_1$ across all teams and game segments.}
    \label{fig:bhatta}
\end{figure}
\noindent
For each team and game segment, we quantify the degree of overlap, denoted as $\mathrm{BC}_1$, in the estimated formation using a single-regime model. Subsequently, we perform the same measurement on the two formations derived from a two-regime model, obtaining coefficients $\mathrm{BC}_{2,1}$ and $\mathrm{BC}_{2,2}$. Figure~\ref{fig:bhatta} illustrates the distribution of the differences $\mathrm{BC}_{2,1} - \mathrm{BC}_1$ and $\mathrm{BC}_{2,2} - \mathrm{BC}_1$. The regimes are ordered such that Regime~1 corresponds to the formation with the lower Bhattacharyya coefficient, i.e., $\mathrm{BC}_{2,1} \leq \mathrm{BC}_{2,2}$. Consistently, Regime~1 exhibits lower overlap values compared to the single-regime formation, whereas Regime~2 shows higher overlap. This indicates that the second regime reliably captures frames with less structured formations, and  the first regime provides a less overlapping formation as a result. These findings highlight the importance of unsupervised regime modeling to identify frames that do not conform to the team’s general structural patterns.

\begin{figure}[htbp]
    \centering
    \begin{minipage}{0.48\linewidth}
        \centering
        \includegraphics[width=\linewidth]{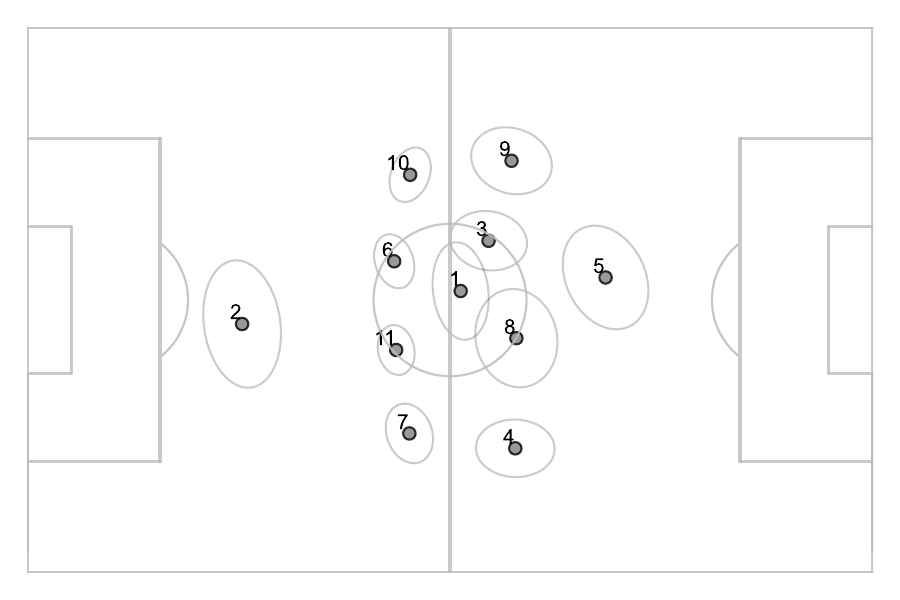}
        \caption*{(a) Regime 1 with probability 31.15\%}
    \end{minipage}\hfill
    \begin{minipage}{0.48\linewidth}
        \centering
        \includegraphics[width=\linewidth]{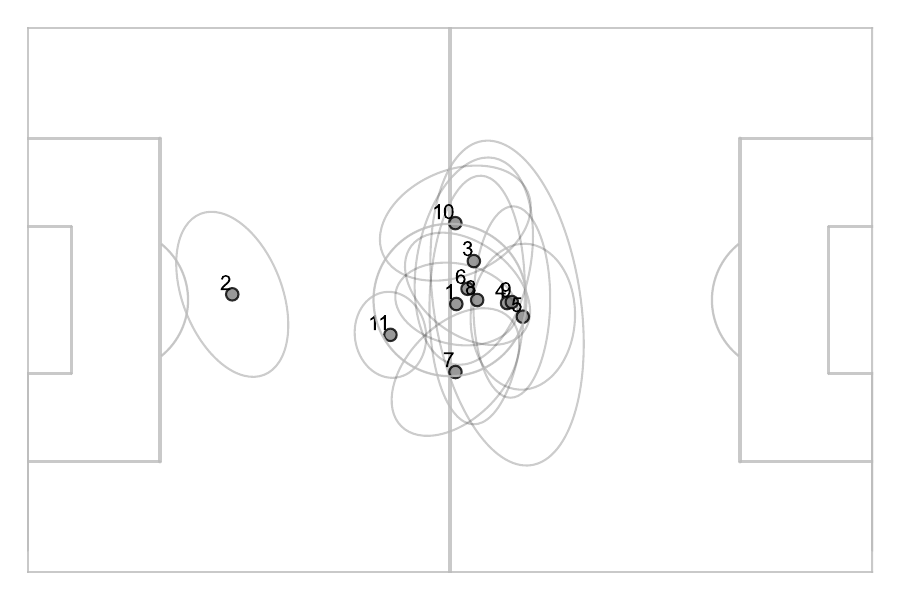}
        \caption*{(b) Regime 2 with probability 15.92\%}
    \end{minipage}
    \begin{minipage}{0.48\linewidth}
        \centering
        \includegraphics[width=\linewidth]{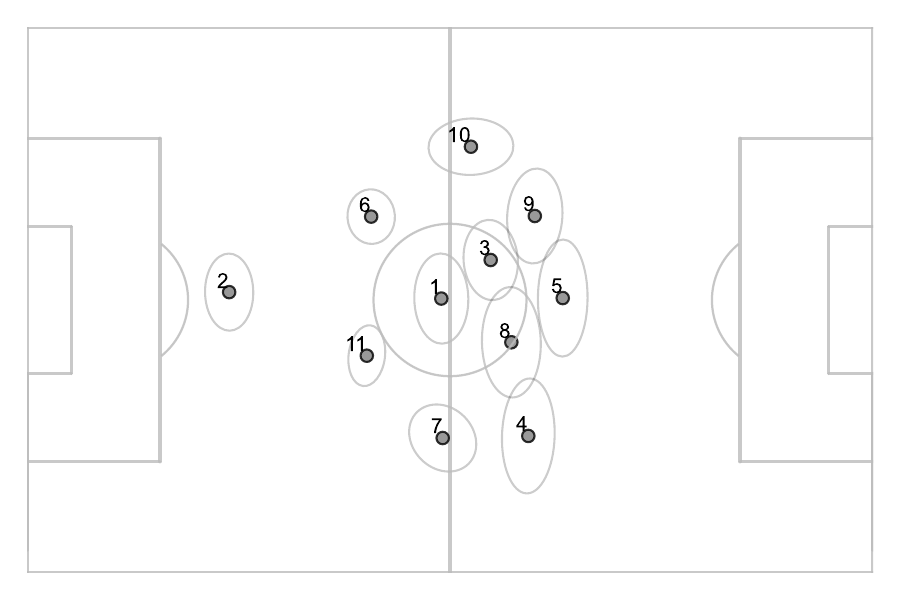}
        \caption*{(c) Regime 3 with probability 51.93\%}
    \end{minipage}
    \caption{Formations estimated using the 3-regime model. Each ellipse corresponds to the empirical covariance of a role position, 
    drawn at one standard deviation around the empirical mean.}
    \label{fig:three_regime_em}
\end{figure}

\begin{figure}
        \centering
        \includegraphics[width=0.7\linewidth]{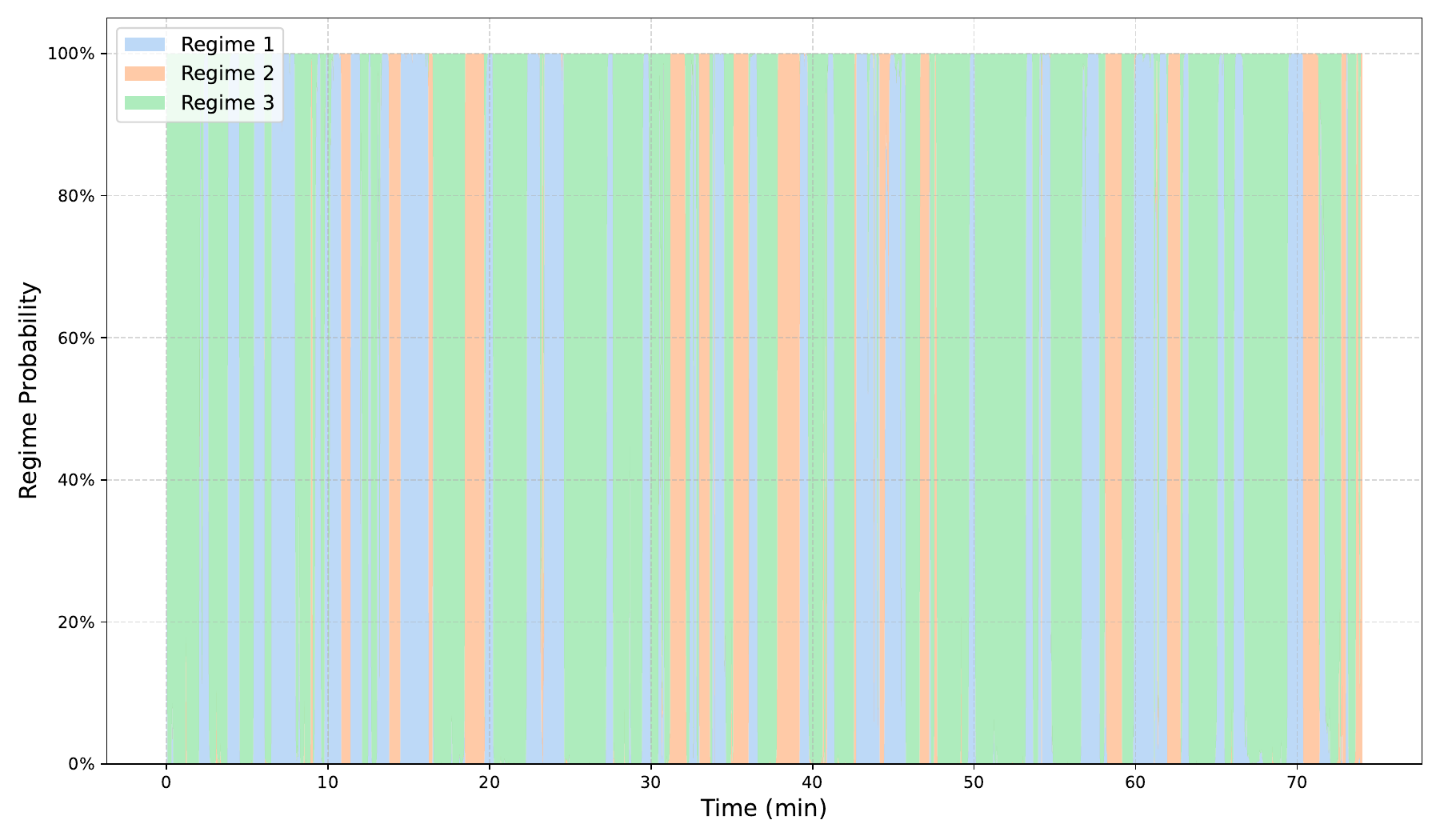}
        \caption{Evolution of regime probability.}
        \label{fig:regimeprobability}
    \end{figure}

\begin{figure}
    \centering
    \includegraphics[width=0.5\linewidth]{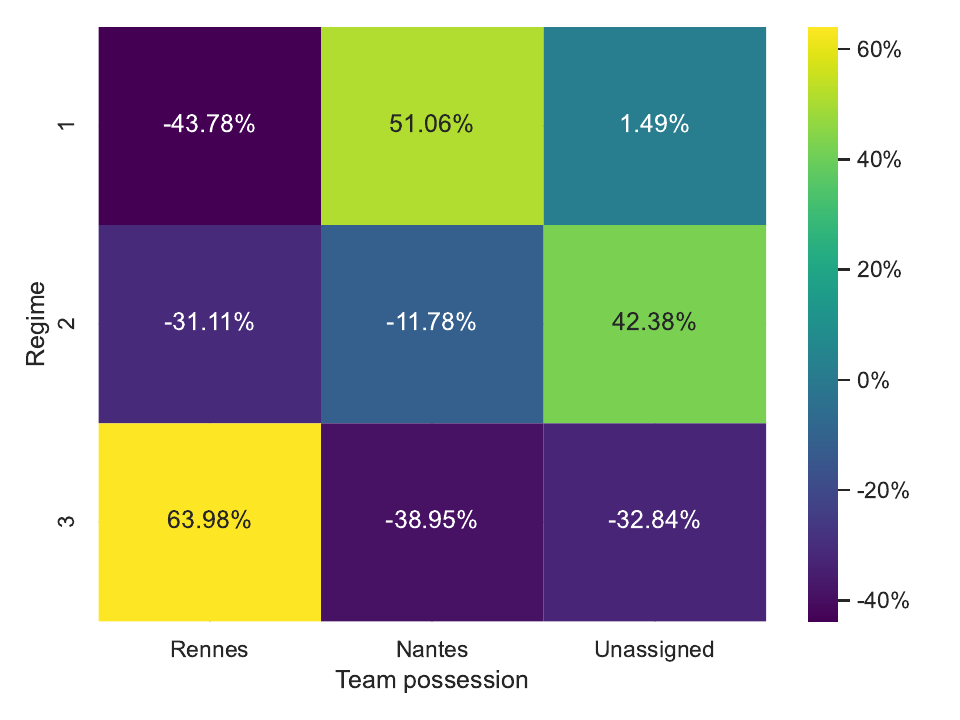}
    \caption{The correlation between the regime probabilities and the possession value across the frames of the game between Rennes and Nantes.}
    \label{fig:corr_poss}
\end{figure}

\paragraph{Three regime model}
We extend this analysis by adding another regime and fitting the model another time. This time, we initialize the mean and covariance parameters of roles by splitting the tracking data into three parts and considering the empirical mean and covariance of each third for each regime. For the permutation weights initialization, we assign an initial weight of $\frac{1}{2}$ to the identity permutation across all regimes and distribute the remaining weight of \(\frac{1}{2}\) equally among the other selected permutations. Figure~\ref{fig:three_regime_em} illustrates the estimated formations in this configuration. Each regime represents distinct tactical sub-states: Regimes~1 and~3 accurately approximate in-possession and out-of-possession formations, respectively, while Regime~2 captures transitional or disorganized situations. Our modeling effort is therefore able to segment the game into phases of play with respect to the spatial distribution in an unsupervised way. In this case, the estimated formations align with those found in the attack/defense dichotomy. 
\\
\noindent
\\
A key strength of the complete mixture model is its ability to infer posterior probabilities of each regime at the frame level. This is achieved using Bayes' formula, as shown in Equation~\eqref{eq:frame_regime_probability}. Figure~\ref{fig:regimeprobability} illustrates the evolution of regime probabilities, highlighting when and how frequently each regime emerges throughout the game. This functionality is particularly valuable for practitioners analyzing the dynamic behavior of a team during a match, as it allows for the identification of specific moments when each formation is adopted or when the team's structure becomes disorganized. In Figure~\ref{fig:corr_poss}, we show the correlation between these posterior regime probabilities and the possession value. The results indicate that Regime~1 is correlated at \(51.06\%\) with Nantes holding the ball, while Regime~3 exhibits a \(63.98\%\) correlation with Rennes possession. These findings are consistent with the formation shapes observed during the manual design of phases of play. Additionally, the second regime shows a positive correlation with unassigned possession, supporting the hypothesis that it captures outlying frames. This observation is not surprising since a team's structure is more likely to be disrupted when neither team has possession.


\begin{figure}
        \centering
        \begin{subfigure}[t]{0.45\linewidth}
            \centering
            \includegraphics[width=\linewidth]{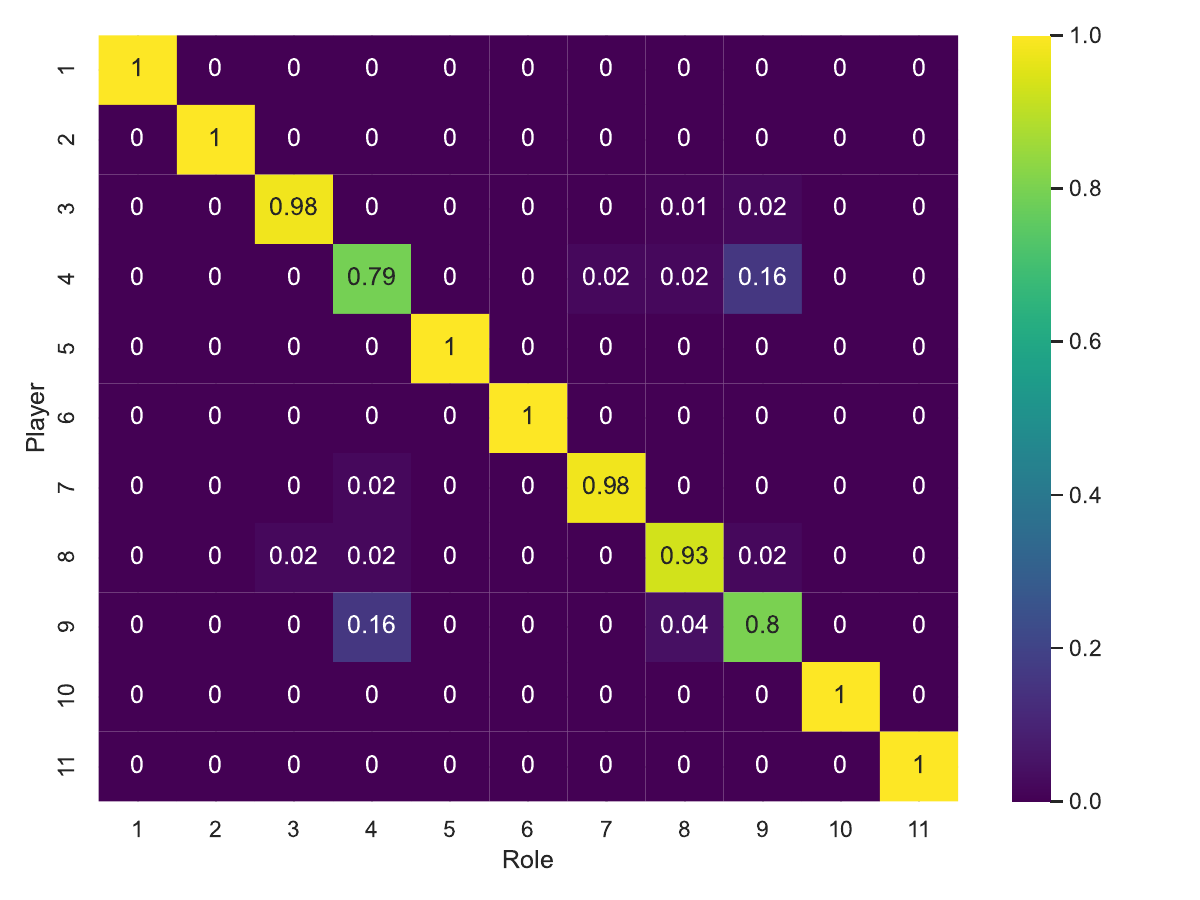}
            \caption{Regime 1 expected role assignment.}
            \label{fig:avg1}
        \end{subfigure}
        \hfill
        \begin{subfigure}[t]{0.45\linewidth}
            \centering
            \includegraphics[width=\linewidth]{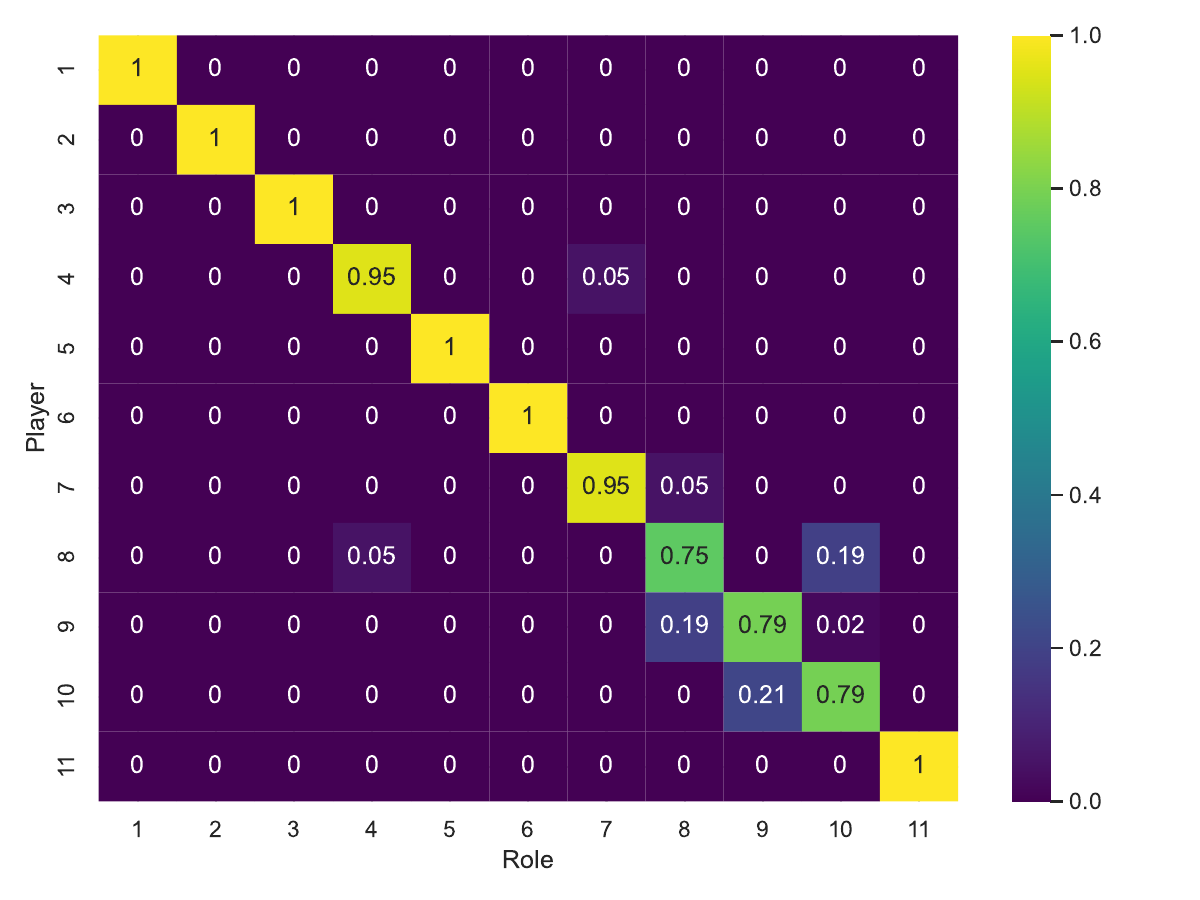}
            \caption{Regime 2 expected role assignment.}
            \label{fig:avg2}
        \end{subfigure}
        \begin{subfigure}[t]{0.45\linewidth}
            \centering
            \includegraphics[width=\linewidth]{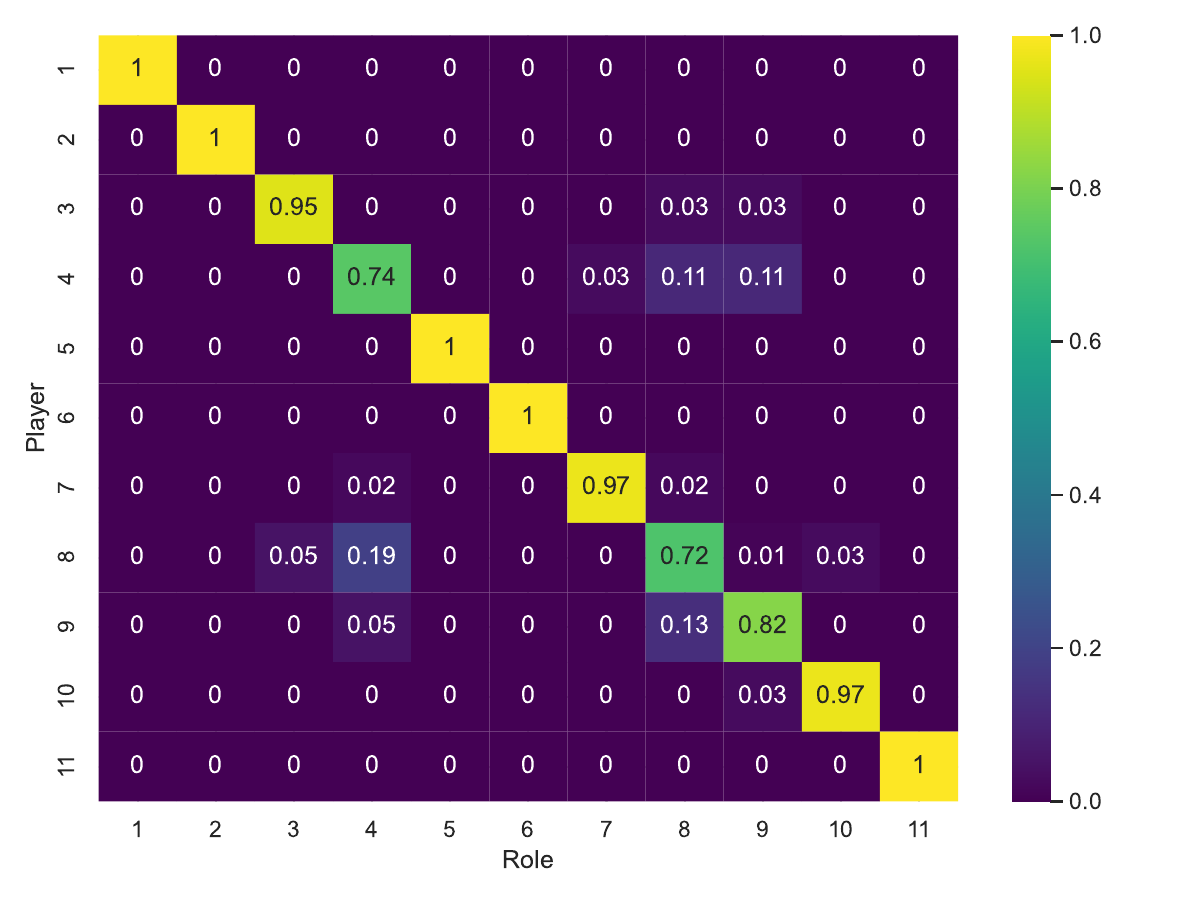}
            \caption{Regime 3 expected role assignment.}
            \label{fig:avg3}
        \end{subfigure}
        \caption{Expected role assignment matrices for each regime in the game of Rennes against Nantes.}
        \label{fig:avgperms}
    \end{figure}

\paragraph{Analysis of the average role assignment}
The model can also be used to track the expected role-to-player assignments within each regime. Figure~\ref{fig:avgperms} illustrates the average permutations in the three regimes, as defined by Equation~\eqref{eq:avg_permutation}. These figures highlight how different role assignments emerge across regimes. Certain roles remain predominantly invariant, exhibiting minimal rates of position swaps. Specifically, the goalkeeper, center-backs, defensive midfielder, and center-forward maintain consistent roles, highlighting the importance of positional rigidity in these critical areas of the pitch. In contrast, wingers demonstrate higher probabilities of role changes. In particular, the two wingers (roles~4 and~9) frequently exchange positions in both offensive and defensive regimes, with a greater rate of swapping during defensive play. This observation is unsurprising, as the side occupied by wingers is less critical when out of possession. Conversely, during offensive play, the wingers' positions are significantly influenced by their dominant foot.

\subsubsection{Probability of no swap}
The analysis in the previous section highlights the advantages of employing the GMM with permutations over the shared component GMM. By integrating hidden permutations into the GMM framework, we are able to derive useful metrics at a granular level using posterior probabilities. More importantly, it enables us to quantify and compute the probabilities of all possible role to player assignments during the game.  This property allows for a more nuanced understanding of team dynamics and tactical rigidity.
\\
\noindent
\\
A key metric that can be derived from this model is the probability of the identity permutation $w_{I_d}$, which serves as an indicator of the rigidity in a team's tactical setup. The identity permutation corresponds to scenarios where players maintain their designated roles without any swaps. A high probability of no swap implies that the team consistently adheres to a predefined tactical formation, reflecting a structured and possibly less adaptable approach. Conversely, a lower probability suggests greater role fluidity, indicating that players frequently interchange positions to adapt to the evolving dynamics of the game.
\\
\noindent
\\
Table~\ref{tab:frame_no_swap} presents the average probability of no swap for each team analyzed. These probabilities are computed by aggregating all game slices for each team across multiple matches, with the averages weighted according to the number of frames in each game slice. The resulting probabilities highlight the varying degrees of role rigidity across different teams. Teams exhibiting higher no-swap probabilities, such as Troyes and Nice, demonstrate a strong adherence to their role assignments throughout the game. This consistency suggests a deliberate strategic approach, potentially limiting the positional freedom of players. On the other hand, teams like Brest and Lorient, with lower no-swap probabilities, display greater role fluidity. This quantitative measure provides valuable insights into the strategic stability and flexibility of each team, enabling a nuanced analysis of their tactical behaviors.

\begin{table}[ht]
\centering
\begin{tabular}{lr}
\toprule
\toprule
Team & Value \\
\midrule
Troyes                 &86.79\% \\
Nice                   &83.55\% \\
Metz                   &83.21\% \\
\midrule
Strasbourg             &80.84\% \\
Angers SCO             &80.79\% \\
Clermont               &80.37\% \\
\midrule
Reims                  &78.19\% \\
Lens                   &77.66\% \\
Olympique Lyonnais     &75.39\% \\
\midrule
PSG                    &74.43\% \\
Monaco                 &74.02\% \\
Olympique Marseille    &73.36\% \\
\midrule
Nantes                 &73.08\% \\
Bordeaux               &71.95\% \\
Lille                  &71.93\% \\
\midrule
Rennes                 &70.94\% \\
Montpellier            &69.54\% \\
Saint-Etienne          &67.57\% \\
\midrule
Brest                  & 61.30\% \\
Lorient                &61.28\% \\
\bottomrule
\bottomrule
\end{tabular}
\caption{Average probability of no player-role swap for each team.}
    \label{tab:frame_no_swap}
\end{table}
 
\subsection{Distance between formations}
\begin{figure}
    \centering
    \includegraphics[width=0.8\linewidth]{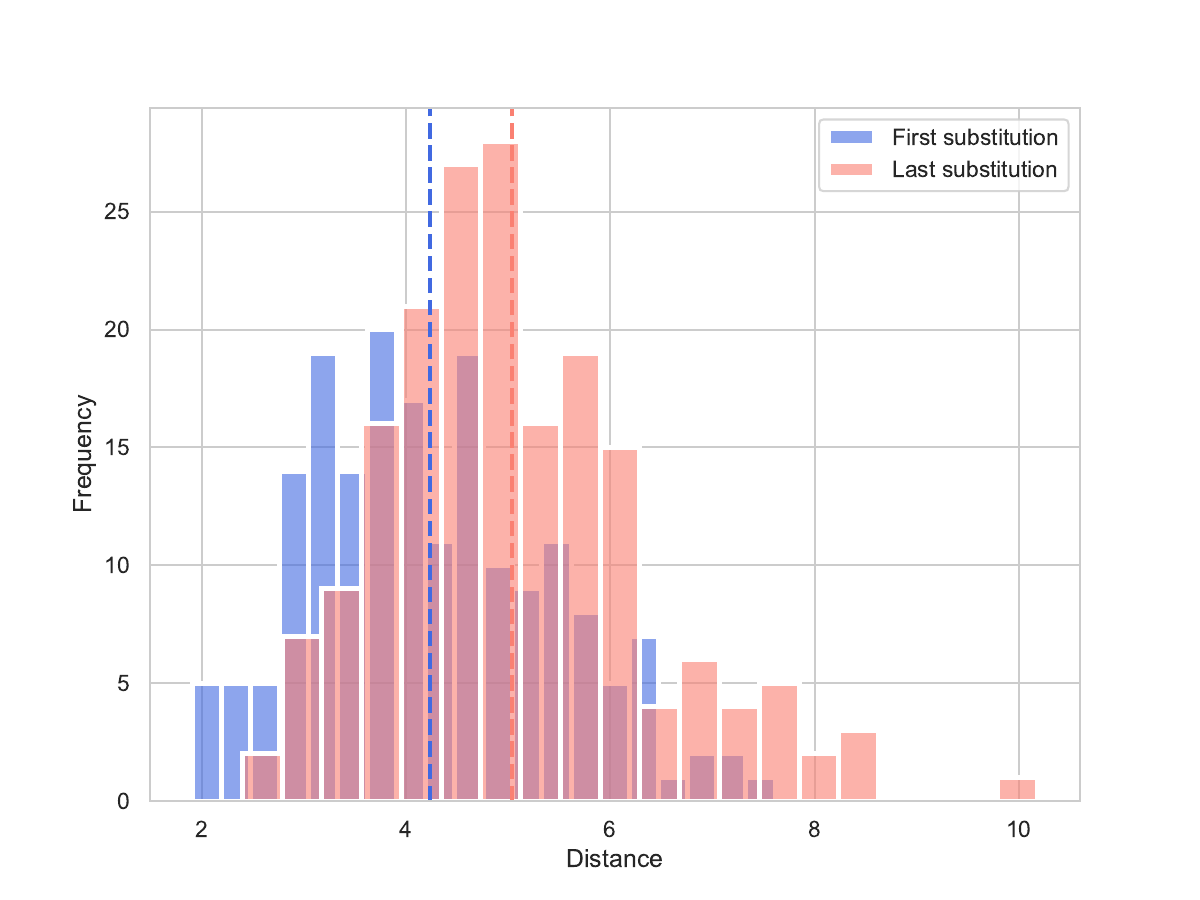}
    \caption{Distribution of the distance between formations during the first substitution and the last substitution.}
    \label{fig:first_last_sub}
\end{figure}

To demonstrate the utility of systematic and unsupervised estimation of formations, we analyze observed game formations using a distance metric between formations. To compare two formations, defined by their estimated role means and covariance matrices, we determine the optimal matching between the roles of each formation to minimize the total distance between roles. The distance between roles used in this analysis is the Wasserstein distance between the corresponding Gaussian distributions, for which a closed-form solution exists. Essentially, this corresponds to the Mixture-Wasserstein distance introduced in \cite{delon2020wasserstein}, where formations are viewed as Gaussian mixtures of their role components.
\\
\noindent
\\
This distance measure can be used to confirm many stylized facts of football tactics. In particular, we investigate the effect of substitutions on the formation. This is done by comparing the formation estimated during consecutive game segments. It should be noted that we only include game segments with at least 5 minutes, and therefore some consecutive game segments can be 5 minutes apart. This can be the case if two substitutions are performed in quick succession. We compare the distance between formations before and after the first substitution of the game, as well as before and after the last substitution.  Figure~\ref{fig:first_last_sub} displays the distribution of these distances. We observe that the last substitution typically results in larger formation changes. This is unsurprising, as the first substitution usually involves a player-for-player change aimed at enhancing performance within the same role. In contrast, the last substitution is often intended to disrupt the opponent's play, either to secure a last-minute goal or to strengthen defensive efforts.
\\
\noindent
\\
These findings confirm that the estimated formations and their distance can capture the strategic shifts that occur after substitutions. This approach provides a versatile tool that can be used for a variety of practical applications for teams. For instance, teams can leverage this distance measure to evaluate the similarity of their structure in a given game to their competitors. 

\subsection{Formation clustering}
\begin{figure}[htbp]
    \centering
    
    \begin{subfigure}[b]{0.45\textwidth}
        \centering
        \includegraphics[width=\textwidth]{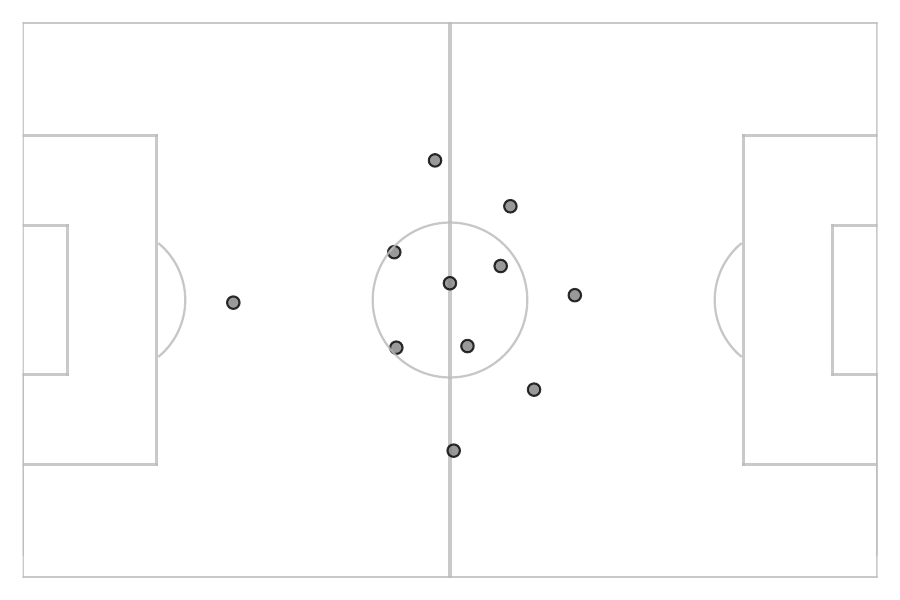}
        \caption{Cluster 1}
        \label{fig:cluster_1}
    \end{subfigure}
    \hfill
    \begin{subfigure}[b]{0.45\textwidth}
        \centering
        \includegraphics[width=\textwidth]{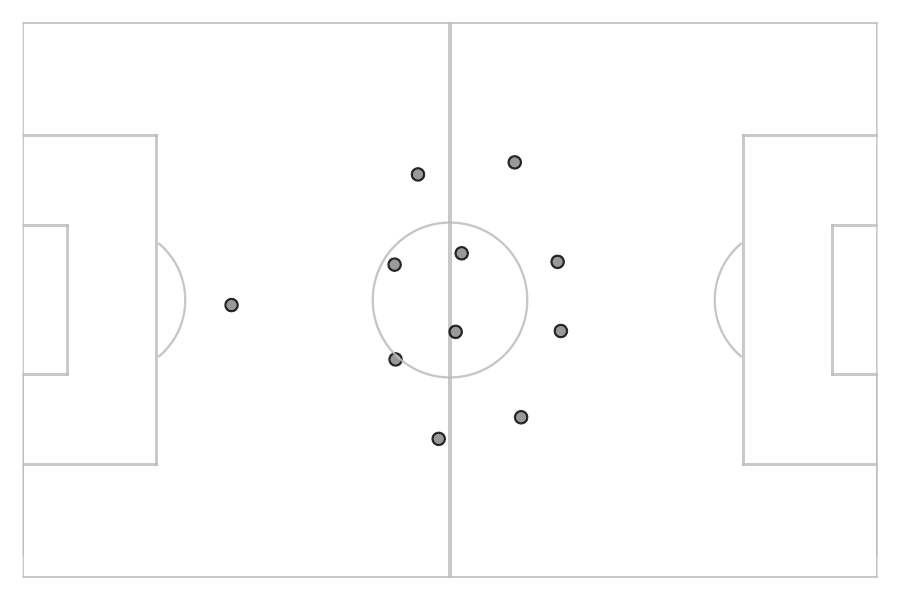}
        \caption{Cluster 2}
        \label{fig:cluster_2}
    \end{subfigure}
    \vspace{0.2cm}
    \begin{subfigure}[b]{0.45\textwidth}
        \centering
        \includegraphics[width=\textwidth]{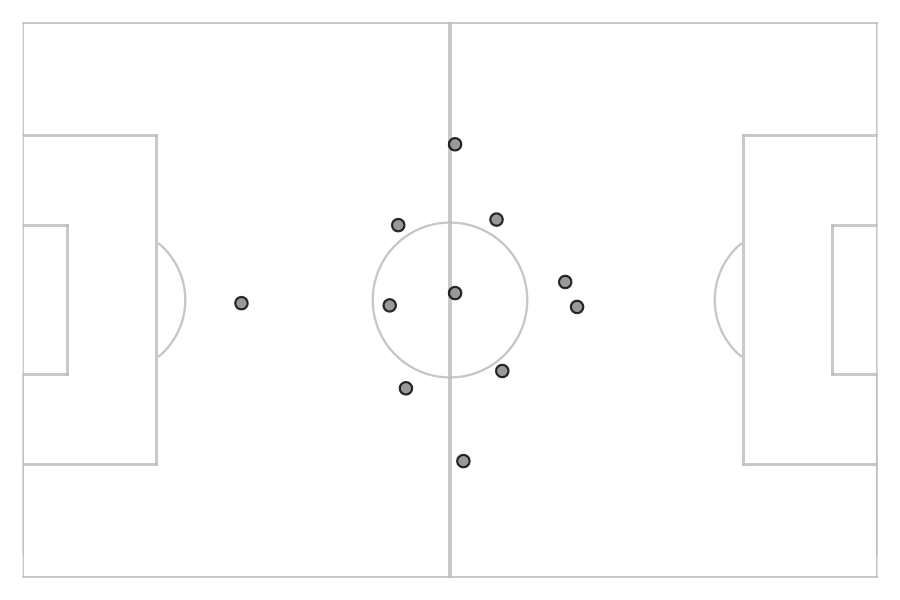}
        \caption{Cluster 3}
        \label{fig:cluster_3}
    \end{subfigure}
    \hfill
    \begin{subfigure}[b]{0.45\textwidth}
        \centering
        \includegraphics[width=\textwidth]{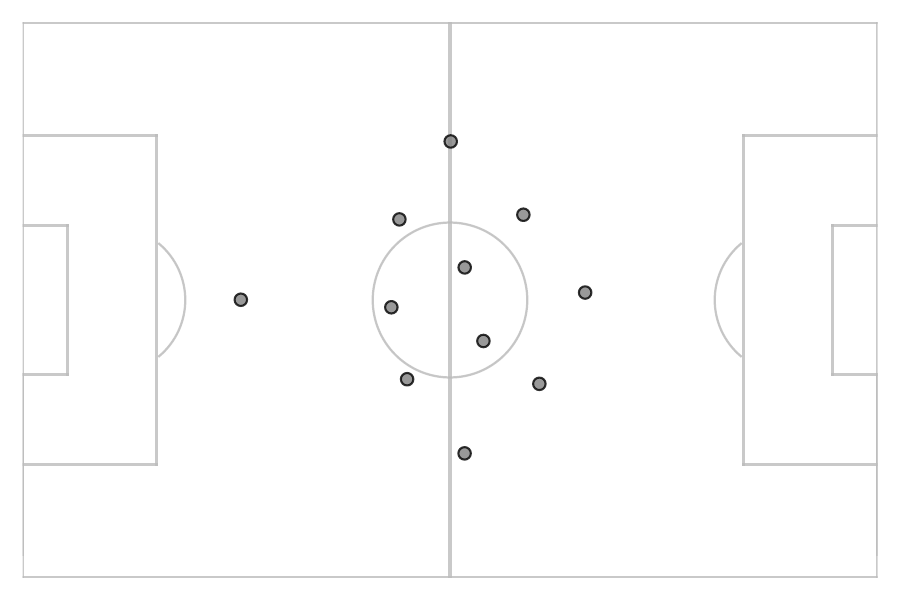}
        \caption{Cluster 4}
        \label{fig:cluster_4}
    \end{subfigure}
    \vspace{0.2cm}
    \begin{subfigure}[b]{0.45\textwidth}
        \centering
        \includegraphics[width=\textwidth]{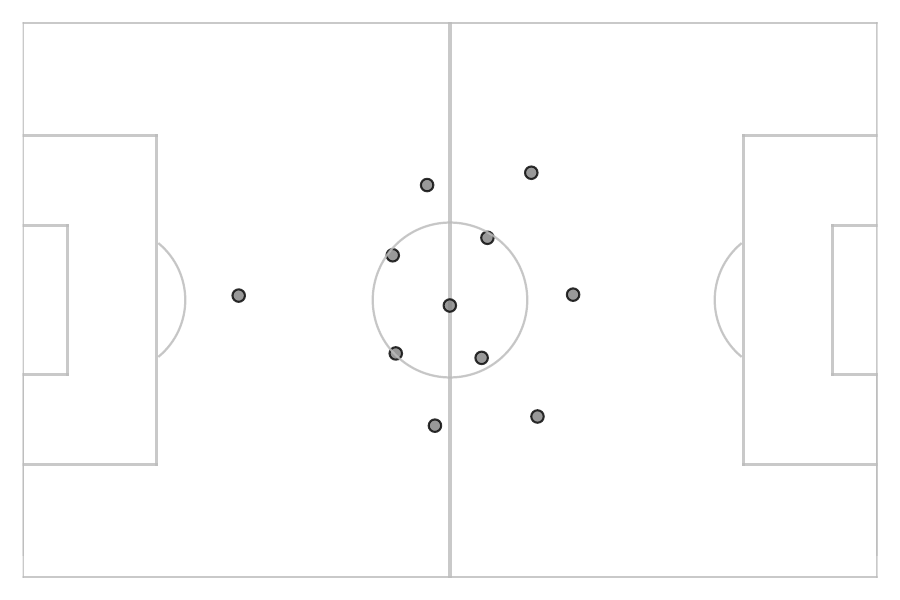}
        \caption{Cluster 5}
        \label{fig:cluster_5}
    \end{subfigure}
    
    \caption{Clusters retrieved using K-means on the embedding of the role centroids.}
    \label{fig:all_cluster_plots}
\end{figure}

\begin{figure}
    \centering
    \includegraphics[width=0.8\linewidth]{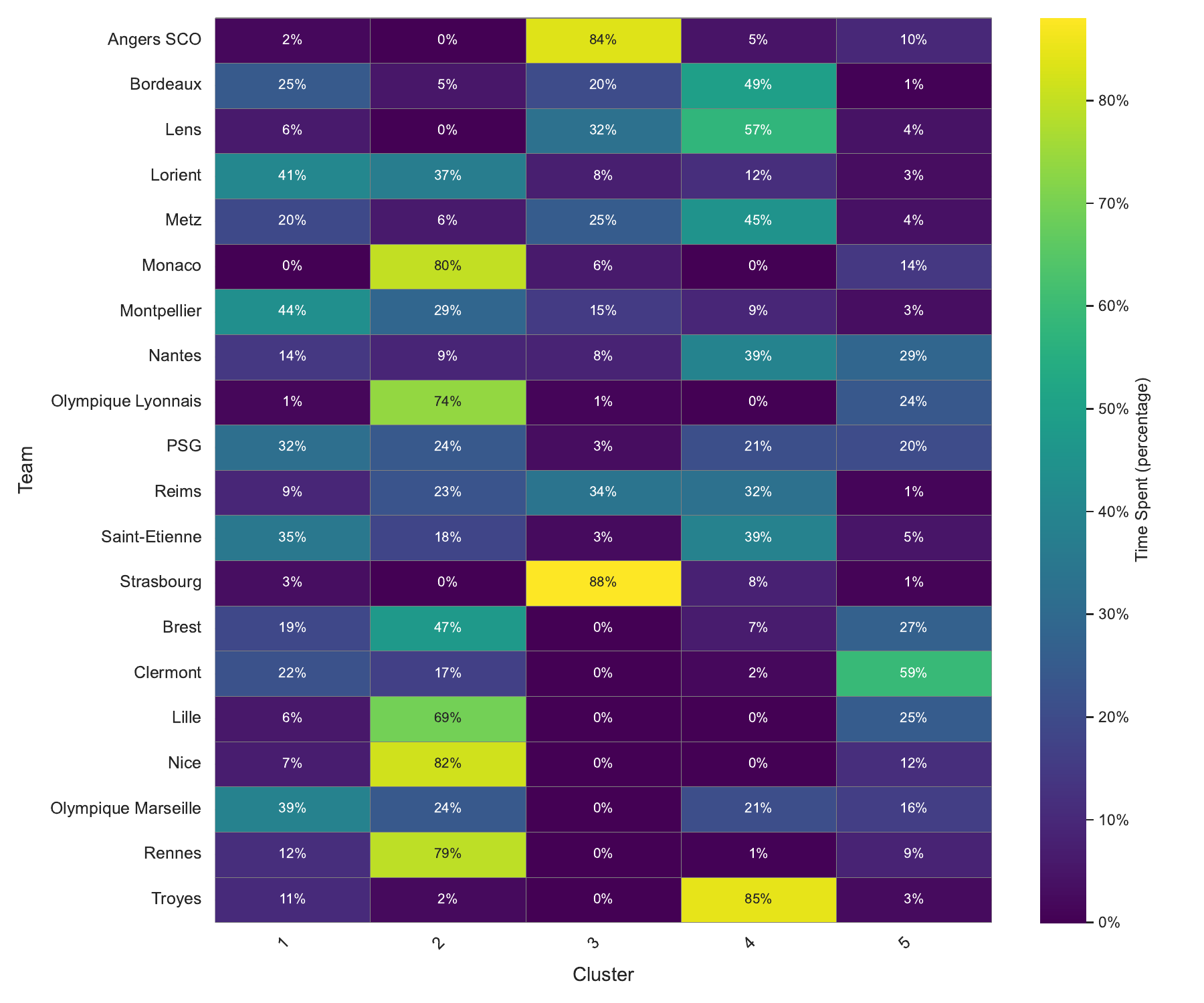}
    \caption{Percentage of time spent in each formation cluster.}
    \label{fig:time_spent_cluster}
\end{figure}

In this section, we aim at grouping similar formations into clusters. We use a different notion of distance between formations to the previous section, to enable the use of the K-means algorithm. For each formation, we keep the mean of each role and project them along $12$ directions in the plane to recover an embedding in $\mathbb{R}^{11\times 12}$, see \cite{baouan2025optimaltransportbasedembedding}. This embedding preserves the sliced-Wasserstein distance and maps formations to Euclidean space, allowing for the efficient use of centroid based algorithms such as K-means. It should be noted that it only takes into account the mean location of roles $\mu$ in the formation and not the shape of the area covered. In Figure~\ref{fig:all_cluster_plots}, each sub-figure displays one of the 5 clusters identified using K-means on the estimated formations of all game slices, capturing patterns that recur across multiple matches. For each cluster, we plot the formation that is closest to the centroid in the embedding space as representative of the cluster.
\\
\noindent
\\
We can recognize well known formations characterized by distinct distributions of player roles: \begin{itemize}
    \item Cluster 1: 4-2-3-1
    \item Cluster 2: 4-4-2
    \item Cluster 3: 5-3-2
    \item Cluster 4: 5-4-1
    \item Cluster 5: 4-3-3
\end{itemize}
We can go further in the analysis and characterize teams with respect to the formation clusters they use the most in Figure~\ref{fig:time_spent_cluster}. This graph can reveal a team’s tactical footprint: Some teams show more diversity in the formations they use across the game segments, whereas others like Strasbourg, Troyes and Angers SCO show stability in the use of their preferred formation cluster. By quantifying the time spent in each cluster, analysts or coaching staff can learn how flexible or rigid a team’s tactics are. Such analysis can also be used by scouting teams to automatically determine the formations a prospective talent is involved in.

%% file: sections/conclusion.tex
\newpage
\section{Conclusion}
In this work, we present a novel probabilistic approach for the automatic detection of formations and the estimation of role assignments in football. The basis of our method is the modeling of the hidden permutation matrix, which allows each player location to be associated with a latent role in a flexible manner. By simultaneously modeling both the distribution of roles and the random assignment of these roles to players, we capture a fundamental feature of modern football: players often swap positions in response to the evolving state of the game. Our framework addresses the cardinality of the set of permutations by introducing parsimonious selection criteria, thus avoiding the otherwise intractable \(d!\) space for \(d=11\) players. As a result, we can systematically identify and retain only those permutations that are statistically significant, effectively encoding the likelihood of key role swaps observed on the pitch.
\\
\noindent
\\
Beyond its applicability for formation detection, our approach is extended to include a latent regime variable, thus extending the analysis to multi-regime scenarios. This addition captures the idea that a team’s formation and role-swapping patterns can vary significantly throughout a match, for instance between in-possession and out-of-possession phases or between standard play and transitional periods. In practice, the multi-regime Gaussian mixture with hidden permutations can separate tactical phases and discover “outlier” frames characterized by disorganized or unusual player arrangements. Such a decomposition is particularly valuable for performance analysts and coaching staff, who benefit from fine-grained insights into how a team’s structure evolves in response to tactical instructions, score-line pressure, or substitutions.
\\
\noindent
\\
Our empirical results on  Ligue 1 tracking data illustrate multiple facets of the method’s utility. We demonstrated that the method can uncover a base formation for a team, reveal how it differs between offensive and defensive situations, and also partition the match into regimes that reflect distinct game states in an unsupervised way. Moreover, the framework enables a collection of derived metrics. We showed how one can compute an index of tactical rigidity, measure formation distances via a Mixture-Wasserstein approach, and cluster formations into commonly recognized patterns for large-scale analysis across multiple matches.
\\
\noindent
\\
From a broader perspective, these contributions have immediate applications in sports analytics and coaching, allowing practitioners to  identify their own and opponents’ formations with minimal manual labeling,  quantify how consistently players occupy certain roles, and detect structural changes across match segments. It opens possibilities for building advanced decision-support tools that integrate with existing match-analysis frameworks. Scouting stands out as another potential application. In particular, measuring the expected role assignment matrix enables us to identify the players who can occupy multiple roles during a game. Finally, the potential extensions of this work are numerous. One immediate avenue is to incorporate temporal dependencies more explicitly, for instance through hidden Markov models that account for the natural correlation of consecutive frames. Another promising direction is to improve the permutation selection mechanism and discard more strictly the unlikely permutations prior to fitting the Gaussian mixture with hidden permutations.

\paragraph{Acknowledgement:} The author thanks Stats Perform, Matthieu Lille-Palette, and Andy Cooper for providing tracking data. He is also grateful to Mathieu Lacome and Sébastien Coustou for inspiring discussions. In addition, the author thanks Sergio Pulido and Mathieu Rosenbaum for their helpful remarks. The author gratefully acknowledges financial support from the chair “Machine Learning \& Systematic Methods in Finance” from Ecole Polytechnique.

%% file: sections/appendix.tex
\newpage
\appendix
\section{Overlap to select permutations}
\label{sec:overlaptoselect}
In this section, we present in detail the procedure to select permutations discussed in Section~\ref{sec:selectingpermutation}. First, we discuss the problem of measuring overlap using empirical samples of two distributions. Then, we present the two classifiers we use in our application. 
\subsection{Measuring the overlap}
\noindent
The degree of overlap between distributions can be quantified by computing the following expression
\begin{equation*}
 v(p,q)=   \int_{\mathbb{R}^d} \min (p(x),q(x))dx,
\end{equation*}
where $p$ and $q$ are the densities of the two distributions. Estimating this overlap coefficient without access to densities poses some difficulty. Therefore, we need a reliable way to approximate it using independent samples. The approach should be scalable since we need to compute it every time we want to discard a permutation. The problem of estimating such integral terms has been studied in the literature. The authors of \cite{poczos2011estimation} provide an estimator based on $k$-nearest neighbor density estimation. In the context of our work, it suffers from a high computational complexity which is prohibitive for large datasets and repeated use. Another approach is the density estimation via histograms, see \cite{freedman1981histogram}; but this method is impacted by the curse of dimensionality, as the number of bins increases exponentially with dimension. It also poses the challenge of the suitable choice of the bin size.
\\
\noindent
\\
To address these challenges, we propose an alternative approach that leverages a classifier to distinguish samples from the two distributions. It is based on the observation that we only need an upper bound for the overlap and not an exact estimate. In fact, we discard a permutation $Q$ if $Y$ and $QY$ show little intersection, and an upper-bound is sufficient for this purpose. In fact, any given classifier's accuracy in distinguishing samples from the two distributions gives a good signal of the level of overlap, as intersecting distributions result in higher error rates.
\\
\noindent
\\
Let $Z_1$ and $Z_2$ be square-integrable random variables in $\mathbb{R}^d$ that admit densities with respect to the Lebesgue measure, denoted by $p$ and $q$, respectively. Let $\Phi$ be a deterministic  classifier, \textit{i.e.} a measurable function defining separate regions of space $\Phi\geq 0$ and $\Phi<0$ to classify realizations of $Z_1$ and $Z_2$.
\begin{proposition}
\label{prop:overlaprate}
The theoretical rate of error of the classifier $a_{\Phi}$ satisfies:
\begin{equation*}
    a_{\Phi}\coloneqq \frac{1}{2}P(\Phi(Z_1)< 0)+\frac{1}{2}P(\Phi(Z_2)\geq 0)\geq \frac{1}{2}v(p,q).
\end{equation*}
\end{proposition}
\noindent
Therefore, $2a_{\Phi}$ provides an upper bound of the overlap between distributions. However, it is a quantity that also needs access to the densities for exact computation. Therefore, we approximate it using the empirical mean:
\begin{equation}
\label{eq:empiricalrate}
\hat{a}_{\Phi}=  \frac{1}{2n} \sum_{i=1}^n \left ( \mathbf{1}_{\Phi(Z_{1,i})< 0}+ \mathbf{1}_{\Phi(Z_{2,i})\geq  0}\right ) .
\end{equation}
Here, $(Z_{1,i},Z_{2,i})$ are i.i.d realizations of $(Z_1,Z_2)$ for $i=1,\dots,n$. Furthermore, the central limit theorem provides confidence intervals on the estimate. We can therefore work with an asymptotic upper-bound with a given level of confidence. 
\\
\noindent
\\
The choice of a suitable $\Phi$ is fundamental to make the quantity $1-\alpha_{\Phi}$ a tight upper-bound. In fact, if chosen arbitrarily the rate of error can be wide and the permutation selection procedure becomes obsolete. To choose an appropriate classifier, some information on the distribution must be incorporated from the data. However, Proposition~\ref{prop:overlaprate} is true for deterministic classifiers that are not constructed using information from any datapoints. In particular, if we plug in the estimator in Equation~\eqref{eq:empiricalrate} a complex enough classifier that is trained using  $Z_{1,i},Z_{2,i}$ for $i=1,\dots,n$, we can achieve lower rates of error. One of two strategies can be used to carefully incorporate information on the distributions:
\begin{enumerate}
    \item Use a simple trained model so that the estimated rate of error converges to the rate of error of a deterministic classifier as the number of data-points increases.
    \item Train the classifier on data-points that are independent from the test data on which the rate of error is calculated.
\end{enumerate}
In this work, we use the two strategies respectively with the two classifiers: Quadratic Discriminant Analysis and Bayesian Gaussian Mixture.
\subsection{Quadratic Discriminant Analysis}
\label{sec:QDA}
Another condition for the choice of the classifier $\Phi$ is that it must be computationally efficient for repeated use. 
\emph{Quadratic Discriminant Analysis (QDA)} is a probabilistic classification algorithm that relies on Bayes' theorem under the assumption that each class follows a Gaussian distribution with its own mean and covariance matrix. While this model makes strong assumptions on the data and is not the most accurate, it is computationally efficient and sufficient for our purposes, which is to discard permutations that show minimal overlap.
\\
\noindent
\\
Let $h$ be the probability density function of a multivariate Gaussian distribution with mean $\mu\in\mathbb{R}^d$ and covariance matrix $\Sigma\in\mathbb{R}^{d\times d}$:
\begin{equation*}
   h(x;\mu,\Sigma) \;=\;
   \frac{1}{\sqrt{(2\pi)^d \, \det(\Sigma)}}
   \exp\Bigl(-\tfrac12\,(x-\mu)^\top\,\Sigma^{-1}\,(x-\mu)\Bigr).
\end{equation*}
\\
\noindent
\\
We define the QDA-based classifier by comparing the likelihoods for two classes:
\begin{equation}
\label{eq:PhiQDA}
   \Phi(y)
   \;=\;
   h\bigl(y;\mu_1,\Sigma_1\bigr)
   \;-\;
   h\bigl(y;\mu_2,\Sigma_2\bigr),
\end{equation}
where $(\mu_1,\Sigma_1)$ and $(\mu_2,\Sigma_2)$ are the mean and covariance of $Z_1$ and $Z_2$, respectively. In practice, these parameters are not known and must be estimated from data. Thus, we replace $(\mu_j,\Sigma_j)$ with their empirical estimates $\bigl(\hat{\mu}_j,\hat{\Sigma}_j\bigr)$ for $j=1,2$, leading to the plug-in estimate
\begin{equation*}
   \hat{a}_{\hat{\Phi}}
   \;=\;
   \frac{1}{2n}\;\sum_{i=1}^n
   \left ( \mathbf{1}_{\hat{\Phi}(Z_{1,i})<0}
   \;+\;
   \mathbf{1}_{\hat{\Phi}(Z_{2,i})\geq 0} \right ),
\end{equation*}
where
\begin{equation*}
   \hat{\Phi}(y)
   \;=\;
   h\bigl(y;\hat{\mu}_1,\hat{\Sigma}_1\bigr)
   \;-\;
   h\bigl(y;\hat{\mu}_2,\hat{\Sigma}_2\bigr).
\end{equation*}
\begin{remark}
The choice of QDA is motivated by the fact that it is parametrized by the mean and covariance only. In the context of our work, we measure the overlap between the distributions of $Y$ and $QY$ to select a permutation matrix $Q$.  Let \( \tilde{Y} \in \mathbb{R}^{2d} \) represent the vectorized version of \( Y \), where for each \( k = 1, \dots, d \), \( \tilde{Y}_{2k-1} = Y_{k,1} \) and \( \tilde{Y}_{2k} = Y_{k,2} \). Similarly, let \( \tilde{Q} \in \mathbb{R}^{2d \times 2d} \) denote the permutation matrix defined by \( \tilde{Q}_{2k-1,2l-1} = \tilde{Q}_{2k,2l} = 1 \)  if \( Q_{k,l} = 1 \) with all the remaining entries equal to zero.  Our objective is to build a classifier that distinguishes samples from \( \tilde{Y} \) and \( \tilde{Q}\tilde{Y} \). To achieve this using QDA, we require the empirical mean and covariance of both vectors. It suffices to compute the empirical mean \( \tilde{\mu} \) and empirical covariance matrix \( \tilde{\Sigma} \) of \( \tilde{Y} \). Consequently, for any permutation \( Q \), the empirical mean and covariance of \( \tilde{Q}\tilde{Y} \) are given directly by \( \tilde{Q}\hat{\mu} \) and \( \tilde{Q}\hat{\Sigma}\tilde{Q}^\top \), respectively. 
\end{remark}
\noindent
Finally, to account for estimation error, we adjust the observed error rate with a confidence term to account for estimation error, yielding an asymptotic upper-bound of the overlap with confidence $1-\alpha$:
\[
  2\hat{a}_{\hat{\Phi}} + \frac{z_{1-\alpha}}{\sqrt{2n}},
\]
where $z_{1-\alpha}$ is the $(1-\alpha)$-th quantile of the standard normal distribution. The following proposition provides statistical justification for this adjustment. 
\begin{proposition}\label{prop:CLT}
    Under the assumption that $(\mu_1,\Sigma_1) \neq (\mu_2,\Sigma_2)$, we have
   \begin{equation*}
      2\sqrt{n}\bigl(\hat{a}_{\hat{\Phi}} - a_\Phi\bigr)
      \;\xrightarrow{\;}\;
      \mathcal{N}\bigl(0,\beta^2\bigr),
   \end{equation*}
   with $\beta^2 \le \tfrac{1}{2}$. Furthermore, 
   $
       2\hat{a}_{\hat{\Phi}} \;+\; \frac{z_{1-\alpha}}{\sqrt{2n}}
   $
   provides an asymptotic upper bound for $v(p,q)$ with confidence $1-\alpha$.
\end{proposition}

\subsection{Bayesian Mixture of Gaussians}
\label{sec:bayesianmixture}
While Quadratic Discriminant Analysis (QDA) offers a straightforward and computationally efficient approach for classification by assuming that each class follows a single Gaussian distribution, this assumption can be restrictive and produce large rates of errors in practice. This, in-turn, can results in upper-bounds of the overlap that are not tight enough and that do not discard a large number of permutations. Specifically, the data points 
 we look to classify $Y$ and $QY$ are both assumed to be mixtures of multiple Gaussian distributions, and QDA may fail to distinguish these classes effectively due to its simplistic modeling. 
\\
\noindent
\\
To address this limitation, we generalize our approach by employing a Bayesian Mixture of Gaussians. Instead of modeling each class with a single Gaussian, we represent each class as a mixture of several Gaussian components. This allows the model to capture multi-modal distributions and better accommodate the inherent variability within each class. Formally, we fit a Gaussian mixture of $K$ components on the i.i.d realizations $Z_{j,i}$ for $j=1,2$ and $i=1,\dots,n$.
\\
\noindent
\\
Once the mixture models for both classes $Z_1$ and $Z_2$ are trained, we employ a Bayesian classifier to estimate the rate of error. 
\begin{equation*}
   \hat{a}_{\hat{\Phi}}
   \;=\;
   \frac{1}{2n}\;\sum_{i=1}^n
   \mathbf{1}_{\hat{\Phi}(Z_{1,i})<0}
   \;+\;
   \mathbf{1}_{\hat{\Phi}(Z_{2,i})\geq 0},
\end{equation*}
with $\hat{\Phi}=\hat{p}_1-\hat{p}_2$ and $\hat{p}_j$ is the density of the Gaussian mixture estimated for $Z_j$ for $j=1,2$.
\\
\noindent
\\
However, this increased flexibility comes with challenges. If the mixtures of Gaussians are trained on the same datapoints used to estimate the rate of error, we lack garantees that the resulting estimation provides an asymptotic upper-bound to the overlap. This is in contrast to QDA, where the model simplicity gives  statistical guarantees because the trained classifier converges to a deterministic one as $n$ grows.  To preserve the upper-bound property on the overlap between distributions in Proposition~\ref{prop:overlaprate}, we adopt the strategy of using different datapoints for training the mixture models to those used to estimate the error rate.
\\
\noindent
\\
In the context of our work, similarly to the QDA approach, once the mixture model parameters are estimated from the samples of the random vector $Y$, they can be efficiently transformed to deduce the parameters of the Gaussian mixture of $QY$ for any permutation $Q$. Given the estimated component weights $w_1,\dots,w_K$ for  \( \tilde{Y} \) and the associated means $\tilde{\mu}_s$ and covariances $\tilde{\Sigma}_s$ for $s=1,\dots,K$, the parameters for $\tilde{Q}\tilde{Y}$ are directly obtained as $\tilde{Q}\Tilde{\mu}_s$ and $\tilde{Q}\tilde{\Sigma}_s \tilde{Q}^\top$ for $s=1,\dots,K$, with the component weights remaining fixed.
\\
\noindent
\\
By extending from QDA to a Bayesian Mixture of Gaussians, we enhance the tightness of the upper-bound estimated to discard permutations. This enables us to prune more aggressively the set of permutations.

\medskip
\noindent

\section{The EM Algorithm}
\label{sec:em_algo}
The Expectation-Maximization (EM) algorithm is a powerful iterative method for finding maximum likelihood estimates when the data involves latent variables, as is the case in our problem. It operates by alternating between an expectation step and a maximization step. The central idea of EM is rooted in the observation that the likelihood function is simpler to calculate and optimize if the latent variables \( R \) and \( \Pi \) were directly observable. In this scenario, the complete-data likelihood is given by:

\begin{equation}
\label{eq:casnonlatent}
\small
\begin{split}
\Tilde{\mathcal{L}}_\theta(y,\mathbf{Q},\mathbf{r}) = & \sum_{i=1}^n \log \left( v_{r_i} w_{r_i,Q_i} \frac{1}{(2\pi)^d \prod\limits_{k=1}^d \sqrt{\det(\Sigma_{r,k})}} \right ) \\
& \quad - \frac{1}{2} \sum_{k=1}^d \left( \mu_{r_i,k} - (Q_i^{\top} y^{(i)})_k \right)^\top \Sigma_{r_i,k}^{-1} \left( \mu_{r_i,k} - (Q_i^{\top} y^{(i)})_k \right ).
\end{split}
\end{equation}
where $\mathbf{Q}=(Q_i)_{i=1,\dots,n}$, $\mathbf{r}=(r_i)_{i=1,\dots,n}$ are observed permutations and regimes. Since these variables are not observed, the EM algorithm maximizes the expectation in Equation~\eqref{eq:casnonlatent} where we average with respect to the posterior distribution of $Q_i$ and $r_i$ conditionally on the observed locations $y_i$. This is done iteratively to update the estimation by alternating between two steps:
\begin{itemize}
    \item \textbf{E-step:} Estimating the posterior distribution of the latent variables \( R \) and \( \Pi \) given the observed data and the current parameter estimates.
    \item \textbf{M-step:} Updating the parameters \( \theta \) by maximizing the expected complete-data log-likelihood based on the posterior distribution obtained in the E-step.
\end{itemize}

\noindent In particular, at iteration \( t \), the EM algorithm computes the parameters \( \theta^{(t+1)} \) as follows:
\begin{equation}
    \label{eq:em_max}
    \theta^{(t+1)} = \arg\max_\theta \mathbb{E}_{\mathbf{R},\mathbf{\Pi} \sim p_{\theta^{(t)}}(.|y)} \left( \Tilde{\mathcal{L}}_\theta(y, \mathbf{\Pi}, \mathbf{R}) \right),
\end{equation}
\noindent
where the posterior distribution \( p_{\theta^{(t)}}(\mathbf{R}, \mathbf{\Pi} \mid y) \) is computed as:

\begin{align*}
    \small p_{\theta^{(t)}}(\Pi^{(i)}=Q, R^{(i)}=r \mid Y^{(i)}=y^{(i)}) 
    &= v_{ir}^{(t)} w_{irQ}^{(t)},
\end{align*}
with \( v_{ir}^{(t)} \) and \( w_{irQ}^{(t)} \) defined  using Bayes formula as follows: 
\begin{align*}
w_{i,r,Q}^{(t)}&=\frac{g\bigl(Q^\top y^{(i)};\mu^{(t)}_r, \Sigma^{(t)}_r \bigr) w_{r,Q}^{(t)}}{\sum\limits_{Q\in \mathcal{P}_d} g\bigl(Q^\top y^{(i)};\mu^{(t)}_r, \Sigma^{(t)}_r \bigr) w_{r,Q}^{(t)}},\\
v_{i,r}^{(t)}&=\frac{\sum\limits_{Q\in \mathcal{P}_d}g\bigl(Q^\top y^{(i)};\mu^{(t)}_r, \Sigma^{(t)}_r \bigr)w_{r,Q}^{(t)}v_r^{(t)}}{\sum\limits_{r=1}^l\sum\limits_{Q\in \mathcal{P}_d}g\bigl(Q^\top y^{(i)};\mu^{(t)}_r, \Sigma^{(t)}_r \bigr) w_{r,Q}^{(t)}v_r^{(t)}}.
\end{align*}
\noindent The objective in Equation \eqref{eq:em_max} can then be rewritten as:
\begin{equation}
\label{eq:objective_rewrite}
\small
\begin{split}
    \theta^{(t+1)} = \arg\max_\theta \Bigg[ \, &
        \sum_{i=1}^n \sum_{r=1}^l \sum_{Q \in \mathcal{P}_d} 
            v_{ir}^{(t)} w_{irQ}^{(t)} \log \left( 
                \frac{v_r w_{r,Q}}{(2\pi)^d \prod\limits_{k=1}^d \sqrt{\det(\Sigma_{r,k})}} 
            \right) \\
        & \quad - \frac{1}{2} v_{ir}^{(t)} w_{irQ}^{(t)}\sum_{k=1}^d 
            (\mu_{r,k} - (Q^{\top} y^{(i)})_k)^\top 
            \Sigma_{r,k}^{-1} 
            (\mu_{r,k} - (Q^{\top} y^{(i)})_k)
    \Bigg].
\end{split}
\end{equation}
\noindent From this reformulation, we can derive the parameter updates:
\begin{align*}
    \mu_{r,k}^{(t+1)} &= \frac{\sum\limits_{i=1}^n \sum\limits_{Q \in \mathcal{P}_d} v_{ir}^{(t)} w_{irQ}^{(t)} (Q^\top y^{(i)})_k}{\sum\limits_{i=1}^n v_{ir}^{(t)}}, \\
    \Sigma_{r,k}^{(t+1)} &= \frac{\sum\limits_{i=1}^n \sum\limits_{Q \in \mathcal{P}_d} v_{ir}^{(t)} w_{irQ}^{(t)} ((Q^\top y^{(i)})_k - \mu_{r,k}^{(t+1)})((Q^\top y^{(i)})_k - \mu_{r,k}^{(t+1)})^\top}{\sum\limits_{i=1}^n v_{ir}^{(t)}}, \\
    v_{r}^{(t+1)} &= \frac{\sum\limits_{i=1}^n v_{ir}^{(t)}}{n}, \\
    w_{r,Q}^{(t+1)} &= \frac{\sum\limits_{i=1}^n v_{ir}^{(t)} w_{irQ}^{(t)}}{\sum\limits_{i=1}^n v_{ir}^{(t)}}.
\end{align*}
\noindent
The update formulas are intuitive, as they adjust the mean and covariance matrix of each role by averaging the positions of all player components. For each role, the contribution of each player in updating its parameters is weighted by the probability of the permutations that map the player to the role. In a hard clustering scheme, players in each frame are assigned exclusively to a single role, whereas we allow for a probabilistic assignment, reflecting uncertainty in role allocation.
\\
\noindent
\\
In this work, the EM algorithm iterations are repeated for a maximum of $200$ steps. The algorithm terminates early if the average log-likelihood value does not change by more than a tolerance of \(10^{-7}\). Furthermore, since the EM algorithm cannot produce weights equal to zero by construction, we prune permutations with weights below $10^{-10}$ at each iteration. This approach not only reduces space and time complexity but also mitigates issues related to floating-point underflow.

\section{Identifiability of the mixture with permutations}
\label{sec:identifiability}
Let \( f(\cdot, \mu, \Sigma, w) \) denote the density function with respect to the Lebesgue measure of \( Y \) as defined in Equation~\eqref{eq:model}, where \( w_Q \) represents the probability associated with permutation matrix \( Q \) for \( Q \in \mathcal{P}_d \). Suppose there exist two sets of parameters, \( (\mu, \Sigma, w) \) and \( (\mu', \Sigma', w') \), such that
\[
f(y, \mu, \Sigma, w) = f(y, \mu', \Sigma', w'),
\]
for almost all $y$ in $\mathbb{R}^{d\times 2}$. A direct consequence of this equality is that all marginal densities must also be identical almost everywhere. For \( l = 1, \dots, d \), the marginal density of \( Y_l \) is given by:
\[
f_{Y_l}(y_l, \mu, \Sigma, w) = \sum_{k=1}^d \pi_{l,k} \mathcal{N}(y_l \mid \mu_k, \Sigma_k),
\]
where \( \pi_{l,k} = \sum_{Q \in \mathcal{P}_d} w_Q Q_{l,k} \) represents the expected value of the \( (l,k) \)-entry in the permutation matrix $\Pi$. Similarly, for the second parameter set, we have:
\[
f_{Y_l}(y_l, \mu', \Sigma', w') = \sum_{k=1}^d \pi'_{l,k} \mathcal{N}(y_l \mid \mu'_k, \Sigma'_k).
\]
Next, we consider the density function \( h(\cdot, \mu, \Sigma, w) \) defined on \( \mathbb{R}^2 \):
\begin{align*}
  \psi(y, \mu, \Sigma, w) &= \frac{1}{d} \sum_{l=1}^d f_{Y_l}(y, \mu, \Sigma, w) \\
  &= \sum_{k=1}^d \frac{1}{d} \mathcal{N}(y \mid \mu_k, \Sigma_k).
\end{align*}
The simplification leverages the fact that the rows of the matrix \( \pi \) sum to 1, as \( \pi \) is the expectation of a permutation matrix. Essentially, \( \psi \) represents the density of a random variable in \( \mathbb{R}^2 \) obtained by uniformly selecting one component of the vector \( Y \). We have $\psi(y, \mu, \Sigma, w)=\psi(y, \mu', \Sigma', w')$ for almost all $y$. By the identifiability theorem for finite mixtures of Gaussian distributions, it follows that there exists a permutation matrix \( P \) such that \( \mu = P\mu' \) and \( \Sigma_k = \Sigma'_{\sigma(k)} \), where \( \sigma \) is the permutation satisfying \( P = (1_{j=\sigma(i)})_{i,j} \), see \cite{yakowitz1968identifiability}.
\\
\noindent
\\
To determine the equality between the weights of permutations, we express the density \( f(\cdot, \mu, \Sigma, w) \) as:
\begin{equation*}
    f(y, \mu, \Sigma, w) = \sum_{Q \in \mathcal{P}_d} w_Q \, g(Q^\top y, \mu, \Sigma),
\end{equation*}
where \( g \) denotes the density function of \( X \) in \( \mathbb{R}^{d \times 2} \). We have:
\begin{align*}
    f(y, \mu', \Sigma', w') &= \sum_{Q \in \mathcal{P}_d} w'_Q \, g(Q^\top y, \mu', \Sigma'),\\
    &=\sum_{Q \in \mathcal{P}_d} w'_Q \, g(P Q^\top y, \mu, \Sigma),\\
    &=\sum_{Q \in \mathcal{P}_d} w'_{QP} \, g( Q^\top y, \mu, \Sigma).
\end{align*}
Using the identifiability of finite mixtures of Gaussian distributions, we deduce that $w'_{QP}=w_Q$ for all permutation matrices $Q$.

\section{Proofs of propositions}
\label{sec:mathproofs}

\paragraph{Proof of Proposition \ref{prop:CLT}.}
We write
\begin{equation}
\label{eq:twoterms2}
   \hat{a}_{\hat{\Phi}}
   \;=\;
   \hat{a}_\Phi
   \;+\;
   \frac{1}{2n}
   \sum_{i=1}^n
   \Bigl[\mathbf{1}_{\hat{\Phi}(Z_{1,i})<0} - \mathbf{1}_{\Phi(Z_{1,i})<0}\Bigr]
   \;+\;
   \Bigl[\mathbf{1}_{\hat{\Phi}(Z_{2,i})\geq 0} - \mathbf{1}_{\Phi(Z_{2,i})\geq 0}\Bigr].
\end{equation}
Let
\begin{equation*}
   \Delta_i
   \;=\;
   \Bigl[\mathbf{1}_{\hat{\Phi}(Z_{1,i})<0} - \mathbf{1}_{\Phi(Z_{1,i})<0}\Bigr]
   \;+\;
   \Bigl[\mathbf{1}_{\hat{\Phi}(Z_{2,i})\geq0} - \mathbf{1}_{\Phi(Z_{2,i})\geq0}\Bigr],
\end{equation*}
where each $\Delta_i$ has the same distribution for $i=1,\dots,n$. By the law of large numbers, $\hat{\mu}_j \to \mu_j$ and $\hat{\Sigma}_j\to \Sigma_j$ almost surely as $n\to \infty$ for $j=1,2$. Consequently, $\hat{\Phi}(Z_{j,1}) \to \Phi(Z_{j,1})$ almost surely. 
\\\noindent\\
Since $(\mu_1,\Sigma_1)\neq (\mu_2,\Sigma_2)$, we have $\Phi(z)\neq 0$  for almost all $z$ in $\mathbb{R}^d$. In fact, $\phi(z)=0$ if and only if 
$$  \log\Bigl(\frac{\det(\Sigma_2)}{\det(\Sigma_1)}\Bigr) -(z-\mu_1)^\top\Sigma_1^{-1}(z-\mu_1) + (z-\mu_2)^\top\Sigma_2^{-1}(z-\mu_2) =0,$$
and the set of roots of a non-zero polynomial has Lebesgue measure equal to $0$. Thus, \(\Phi(Z_{j,1})\neq 0\) almost surely for \(j=1,2\). It follows that \(\Delta_1 \to 0\) almost surely, and since \(\Delta_1\) is bounded, \(\mathbb{E}[\Delta_1^2] \to 0\). By Cauchy--Schwarz,
\begin{align*}
   \mathbb{E} \left [\!\Bigl(\frac{1}{2n}\sum_{i=1}^n\Delta_i\Bigr)^2\right ]
   &\;\le\;
   \frac{n}{4n^2} \;\mathbb{E}[\sum_{i=1}^n\Delta_i^{2}]
   \;=\;
   \tfrac14\mathbb{E}[\Delta_1^{2}]
   \;\longrightarrow\; 0.
\end{align*}
Hence, the second term in \eqref{eq:twoterms2} converges to $0$ in $L^2$. For the first term, by the central limit theorem, we have
\[
   2\sqrt{n}\bigl(\hat{a}_\Phi - a_\Phi\bigr)
   \;\xrightarrow{\;}
   \mathcal{N}(0,\beta^2),
\]
where $\beta^2
   \;=\;
   \mathrm{Var}\bigl[\mathbf{1}_{\Phi(Z_{1,1})\ge0} + \mathbf{1}_{\Phi(Z_{2,1})<0}\bigr]
   \;\le\;
   \tfrac12.$
By Slutsky's lemma,
\[
   2\sqrt{n}\bigl(\hat{a}_{\hat{\Phi}} - a_\Phi\bigr)
   \;\xrightarrow{\;}
   \mathcal{N}(0,\beta^2).
\]
Thus,
\[
   \lim\inf_{n} P\Bigl(v(p,q) \;\le\; 2\hat{a}_\Phi + \tfrac{z_{1-\alpha}}{\sqrt{2n}}\Bigr)
   \geq 
   1-\alpha,
\]
which shows that
$2\hat{a}_{\hat{\Phi}} + \tfrac{z_{1-\alpha}}{\sqrt{2n}}$
serves as an asymptotic upper bound for $v(p,q)$ with confidence $1-\alpha$.
\paragraph{Proof of Proposition~\ref{lemma:overlap_permutation}}
The first inequality is a consequence of the fact that $Y=X$ conditionally on $\Pi=I_d$ and $Y=QX$ conditionally on $\Pi=Q$. Likewise, we have $QY=QX$ conditionally on $\Pi=I_d$ and $Y=X$ conditionally on $\Pi=Q^\top$. To establish Equation~\eqref{eq:overlap_Q}, we write:
 \begin{align*}
p(y)&\geq P(\Pi=I_d) f_X(y) + P(\Pi=Q) f_{QX}(y),\\
&\geq \min(P(\Pi=I_d),P(\Pi=Q^\top))f_X(y)+\min(P(\Pi=Q),P(\Pi=I_d))f_{QX}(y).
\end{align*}
And similarly :
\begin{align*}
q(y)&\geq P(\Pi=Q^\top) f_X(y) + P(\Pi=I_d) f_{QX}(s),\\
&\geq \min(P(\Pi=I_d),P(\Pi=Q^\top))f_X(y)+\min(P(\Pi=Q),P(\Pi=I_d))f_{QX}(y).
\end{align*}
By integrating $\min(p,q)$, we can deduce that: 
\begin{align*}
    v(p,q)\geq   \min(P(\Pi=I_d),P(\Pi=Q^\top))+\min(P(\Pi=Q),P(\Pi=I_d)).
\end{align*}